\documentclass[12pt,preprintnumbers,aps,amsmath,onecolumn,nofootinbib]{revtex4-1}

\usepackage{graphicx}
\usepackage{hyperref}
\usepackage{lineno}

\begin{document}

\title{\Huge  Advances in Quark Gluon Plasma}

\author{Gin\'es Mart\'inez Garc\'ia}

\affiliation{SUBATECH, CNRS/IN2P3, \'Ecole des Mines de Nantes, Universit\'e de Nantes, 4 rue Alfred Kastler, 44307 Nantes, France}

\date{\today}

\begin{abstract} 
 In the last 20 years, heavy-ion collisions have been a unique way to study the hadronic matter in the laboratory. 
Its phase diagram remains unknown, although many experimental and theoretical studies have been undertaken in the last decades.
After the initial experiences accelerating heavy nuclei onto fixed targets at the AGS (BNL, USA) and the SPS  (CERN, Switzerland), the \emph{Relativistic Heavy Ion Collider} (RHIC) at BNL was the first ever built heavy-ion collider. RHIC delivered its first collisions in June 2000 boosting the heavy-ion community.
Impressive amount of experimental results has been obtained by the four major experiments at RHIC: PHENIX, STAR, PHOBOS and BRAHMS. In November 2010, the \emph{Large Hadron Collider} (LHC) at CERN delivered lead-lead collisions at unprecedented center-of-mass energies, 14 times larger than that at RHIC. The three major experiments, ALICE, ATLAS and CMS, have already obtained many intriguing results. Needless to say that the heavy-ion programs at RHIC and LHC promise fascinating and exciting results in the next decade.

The first part of the lectures will be devoted to introduce briefly the QCD description of the strong interaction (as part of the Standard Model of Particle Physics) and to remind some basic concepts on phase transitions and on the phase diagram of matter. 

In the second part, I will focus on the properties of matter at energy densities above $\approx$1 GeV/fm$^3$. A historical approach will be adopted, starting with the notion of limiting temperature of  matter introduced by Hagedorn in the 60's and the discovery of the QCD asymptotic freedom in the 70's. The role played by the chiral symmetry breaking and restoration in the QCD phase transition will be discussed, supported by an analogy with the ferromagnetic transition. The phase diagram of hadronic matter, conceived as nowadays, will be shown together with the most important predictions of lattice QCD calculations at finite temperature. Finally, the properties of an \emph{academic} non-interacting ultra-relativistic QGP and its thermal radiation will be deduced.  The dissociation of the heavy quarkonium due to the color-screening of the heavy-quark potential will be described, based on a QED analogy. The energy-loss phenomenology of ideal long-living partons traversing the QGP, will be reminded.

In the third part, the heavy-ion collisions at ultra-relativistic energies will be proposed as a unique experimental method to study QGP in the laboratory, as suggested by the Bjorken model. The main experimental facilities in the world will be described, namely the CERN and BNL accelerator complexes. The main probes for characterizing the QGP in heavy-ion experiments, followed by a brief description of the main heavy-ion experiments located at these facilities will be shown.

In the last part of these lectures, I will present  my \emph{biased} review of the numerous experimental results obtained in the last decade at RHIC which lead to the concept of strong interacting QGP, and the first results obtained at LHC with  the 2010 and 2011 PbPb runs. Finally, the last section is devoted to refer to other lectures about quark gluon plasma and heavy ion physics.

\end{abstract}

\maketitle

\tableofcontents

\newpage 

\section{Introduction}

The strong interaction, described by quantum chromodynamics (QCD), is the dominant interaction in the subatomic world. The main properties of the strong interaction are:
\begin{itemize} 
\item the strength constant $\alpha_s$ at low energies is large, and as a consequence, quantitative calculations based on a perturbative sum of Feynman diagrams, fail\footnote{Roughly, the strength constant $\alpha_s$ can be estimated from the hadron bound state properties. Using the Bohr radius expression of the hydrogen atom (please forgive me this unacceptable assumption) $r_B=2/(m\alpha)$, one can estimate $\alpha_s\sim20$, considering $r_B=~1~{\rm fm}$ and $m_q=~10~{\rm MeV}$.};
\item gluons, $g$ (the intermediate boson of strong interaction) are coloured (colour is the charge of the strong interaction). For this reason, the QCD becomes a complex quantum field theory (QFT), which belongs to the class of non-abelian QFT.
\end{itemize}

The quarks $u$, $d$ and $s$, also called light quarks, exhibit small masses and therefore the most important parameter of QCD is indeed $\alpha_s$. However $\alpha_s$ can only be determined in the high energy domain, since its experimental determination at low energy is difficult due to non-perturbative effects. One of the major experimental observations that  QCD should explain, is the confinement of quarks and gluons. Coloured free particles do not exist, and thus quarks and gluons seem to be confined inside colourless particles called hadrons. The confinement property is not fully understood, despite the fact that the quark model describes qualitatively the hadron properties (mesons are bound states of a quark and antiquark and baryons are bound states of 3 quarks). Today, the best \emph{ab-initio} quantitative calculations can be performed via lattice calculations of QCD. One should note that the origin of the hadron mass is the strong interaction, since light quark masses only represent less than 10\% of the total hadron mass. As  Frank Wilczek (Nobel prize in 2004) expressed in  Physics Today in November 1999   \emph{"According to quantum chromo-dynamics field theory, it is precisely its color field energy that mostly make us weigth.  It thus provides, quite literally,  mass without mass".} In this respect the Higgs boson (strictly speaking the Brout-Englert-Higgs boson or BEH boson) only explains about 1\% of the total mass of the proton and neutron which are the main massive constituents of ordinary matter.

\subsection{Asymptotic freedom}
\begin{figure}
{\centering 
\resizebox*{0.45\columnwidth}{!}{\includegraphics{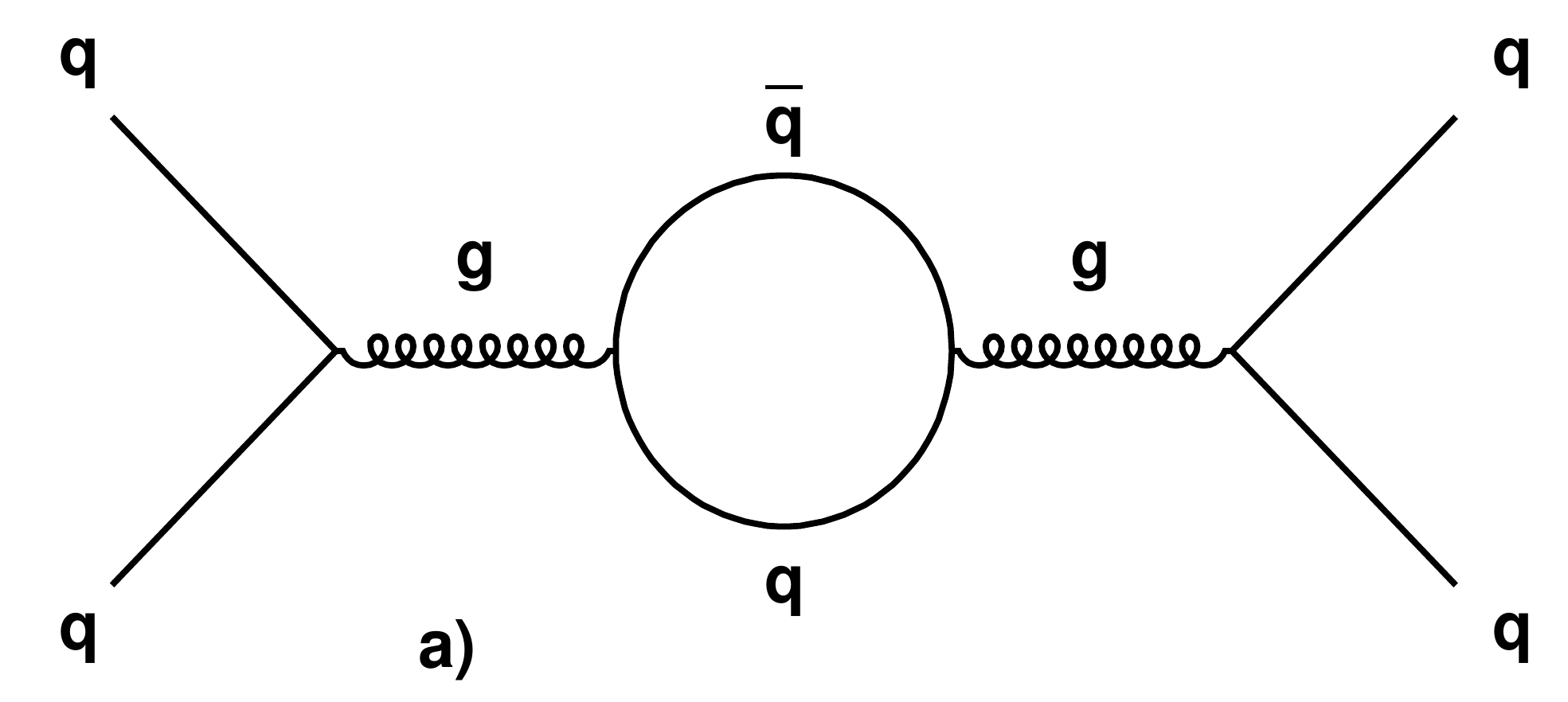}}
\resizebox*{0.45\columnwidth}{!}{\includegraphics{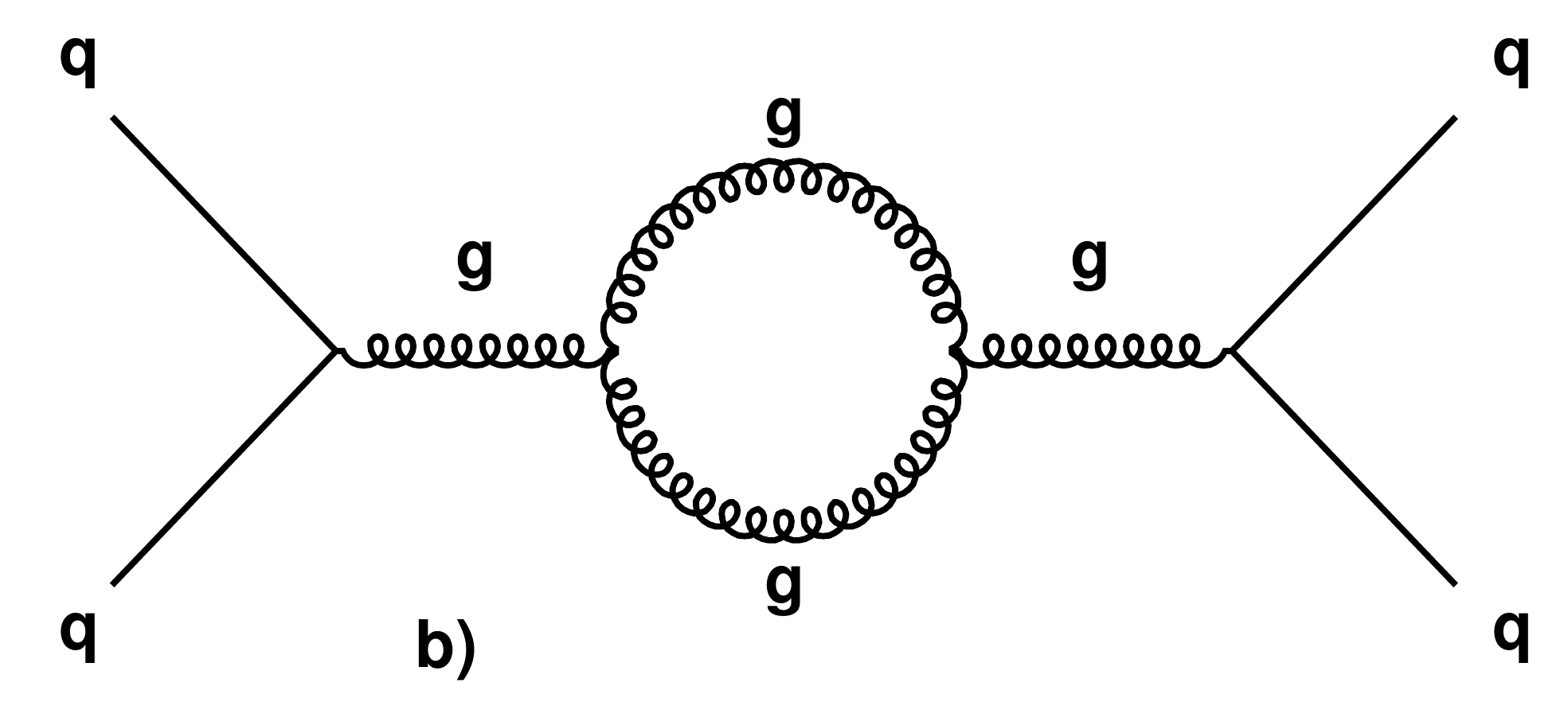}}
\par}
\caption{\label{Intr_PolaVide}
\emph{Feynman diagrams at first order, of the vacuum polarization in QCD: a) screening and b) anti-screening. In the case of QED, anti-screening does not exist since photons are not charged particles.}}
\end{figure}

The vacuum polarization of QCD  \cite{Poli73, Gros73} exhibits a singular  behaviour  due to the anti-screening effect  of virtual gluon pair production (remember that gluons are colored  bosons and the gluon vertex does exist in QCD). Indeed the gluon anti-screening is stronger that the screening effect of virtual quark pair production (see Fig. \ref{Intr_PolaVide}). In QCD one gets  \cite{Grif87}:
\begin{equation}
\alpha(|q^2|) = \frac{\alpha_s(\mu^2)}{1+\frac{\alpha_s(\mu^2)}{12\pi}(11n-2f)\ln{(|q^2|/\mu^2)} }
\label{lib_asym}
\end{equation}
where $n$ represent the number of colors and $f$ the number of quark flavors. In nature, $11n>2f$, in consequence, the strength of the strong interaction $\alpha_s$ decreases at small distances (or high energies) . This phenomenon is called the \emph{asymptotic freedom} of QCD. The discovery of the asymptotic freedom was awarded with the \href{http://www.nobelprize.org/nobel_prizes/physics/laureates/2004/}{Nobel Prize in 2004\footnote{\url{http://www.nobelprize.org/nobel_prizes/physics/laureates/2004/}}}.

At the scale of the Z boson mass ($|q^2|\sim M_Z$), $\alpha_s$ has been measured via many different physics channels and the current world average is $0.1184\pm0.0007$ \cite{PDG12}. Perturbative QCD calculations are then fully valid to describe the strong interaction at  high energies. This is one of the major successes of QCD theory.

In equation (\ref{lib_asym}), the parameter $\mu^2$ is imperative since perturbative QCD calculations cannot be performed in the domain where $\alpha_s\gtrsim1$  (when $|q^2| \lesssim 1$ GeV$^2$). The expression (\ref{lib_asym}) can be rewritten as a function of a parameter $\Lambda_{QCD}$:
\begin{equation}
\alpha(|q^2|) = \frac{12\pi}{(11n-2f)\ln{(|q^2|/\Lambda_{QCD}^2)} }.
\label{lib_asym2}
\end{equation}
 $\Lambda_{QCD}$ is $213\pm8$ MeV for 5 flavours  \cite{PDG12}. However, the quantity $\Lambda_{QCD}$ is not well defined. Therefore it has become standard practice to quote the value of $\alpha_s$ at a given scale (typically $M_Z$) rather than to quote a value for $\Lambda_{QCD}$  \cite{PDG12}.

\subsection{Lattice QCD calculations}
As we have previously seen, the asymptotic freedom is the main success of QCD and it allows for the experimental test of QCD via the study of high energy processes. I would say that the description of  i) the evolution of the parton distribution functions at low Bjorken x values, ii) the production of jets in elementary collisions and iii) the properties of bottomonium bound states , represent beautiful examples of the success of QCD predictions in its perturbative domain.

Nevertheless QCD should a priori explain many other phenomena where perturbative calculations are forbidden. Fundamental questions such as the coupling constant value, QCD vacuum structure, hadron masses, hadron structure, nuclear properties etc ... should be explained by the QCD theory. Even today, the test of QCD has not been possible in many of these domains.  In order to palliate this, many effective models have been developed in the domain of hadronic and nuclear physics. 

Lattice calculation for gauge theories is the most promising non-perturbative technique to solve QCD equations.  The space-time continuum is discretized in a finite number of points where the equations of the theory can be solved.  In the last decade, the impressive development of computing hardware and the optimization of software algorithms have allowed lattice QCD calculations to become a competitive tool \cite{Wilc03}. Nowadays computing facilities for lattice QCD calculations are a crucial component of this research, at the same level as accelerators, detectors or computing centers for data analysis and data storage. Today lattice QCD provides the most precise computation of the $\alpha_s$ constant, is able to extract the mass of the quarks and to predict the mass of most of the hadrons, and it makes excellent predictions of the exotic structure of new bound states of heavy quarks.  In particular, we will see later in more detail,  how lattice QCD at finite temperature allows  to study the hadronic matter phase diagram.

\subsection{A description of the hadronic matter phase diagram} 

By hadronic matter I mean that in which the strong interaction is the main interaction between elementary constituents, that provide the proper degrees of freedom of the matter.
At temperatures above $10^9$ K (1 MeV) and/or pressures above 10$^{32}$ Pa (1 MeV/fm$^3$), the strong interaction is expected to be the dominant interaction between the constituents of matter. 
At low temperatures and  a pressure above 1 MeV/fm$^3$ the matter can be described as a degenerated gas of neutrons\footnote{Actually, it is a degenerate gas of baryons and electrons which is a more stable system that a pure neutron gas.}. Such a state, which is very close to the atomic nucleus structure, should exist in the neutron stars. In these stellar objects, a mass slightly larger that the sun mass, is confined in a ten kilometer radius sphere, and densities as high as 10$^{17}$ kg/m$^3$ are reached. 
For higher pressures, above 10$^{35}$ Pa (1 GeV/fm$^3$), the repulsive force of the degenerate gas of neutrons cannot compensate the pressure, and matter is expected to become a low temperature gas of quarks which are not any more confined inside hadrons.  In this exotic state of matter, quark-quark Cooper pairs might exist creating a kind of color superconductor matter \cite{Raja00}. 
On the other hand, the neutron matter should become a gas of nucleons, if it is heated to temperatures of several MeV. Indeed  the nucleon-nucleon potential has some similitudes with the Van der Waals force between molecules. For this reason, it is expected that the neutron matter evaporates into a gas of nucleons at a temperature of about 10 MeV\footnote{Naively, this value can be accepted  since the bound energy of nucleus in the nuclei saturates to a value of 8 MeV.  The liquid-gas transition has been studied in heavy ion collisions at intermediate energies, although it is not discussed in these lectures.}, like the liquid-gas phase transition in ordinary matter.

At very high temperatures and pressures, the nucleon gas (that has become a hadron gas at temperature above 100 MeV) could go through a transition to a deconfined state of matter. This is expected due to the vacuum polarization at the origin of  the asymptotic freedom of QCD. Therefore the strength of the strong force decreases at high temperature. The deconfined state of matter, in analogy with the electromagnetic plasma where ions and electrons are dissociated, has been called Quark Gluon Plasma (QGP)\footnote{Professor E. Shuryak  proposed the name Quark Gluon Plasma in the 80's \cite{Shu78}.}. The transition to QGP takes place at temperatures about 200 MeV($\sim$2$\cdot$10$^{12}$ K), when quarks and gluons are not confined in colorless particles and they become the pertinent degrees of freedom of the system. Other properties of QCD also predict the reason that a phase transition should occur at high temperature.  In quantum field theories, the symmetries of the Lagrangian can be spontaneously broken at low energies or temperatures. In the case of QCD, the spontaneous breaking of the chiral symmetry takes place at low temperature. The restoration of the chiral symmetry at high temperatures becomes a sufficient condition for the existence of a phase transition \cite{Smil03}.

Finally, for temperatures above 10$^{16}$ K,  it is hard to know what would be the structure of matter. Some authors have speculated about new phenomena like formation of microscopic black holes or unification of interactions, etc ... that could appear. Exotic ideas like the formation of a superstring gas have been proposed for temperatures of 10$^{32}$~K \cite{Bowi85}.

\begin{figure}
{\centering
{\resizebox*{1.0\columnwidth}{!}{\rotatebox{0}{\includegraphics{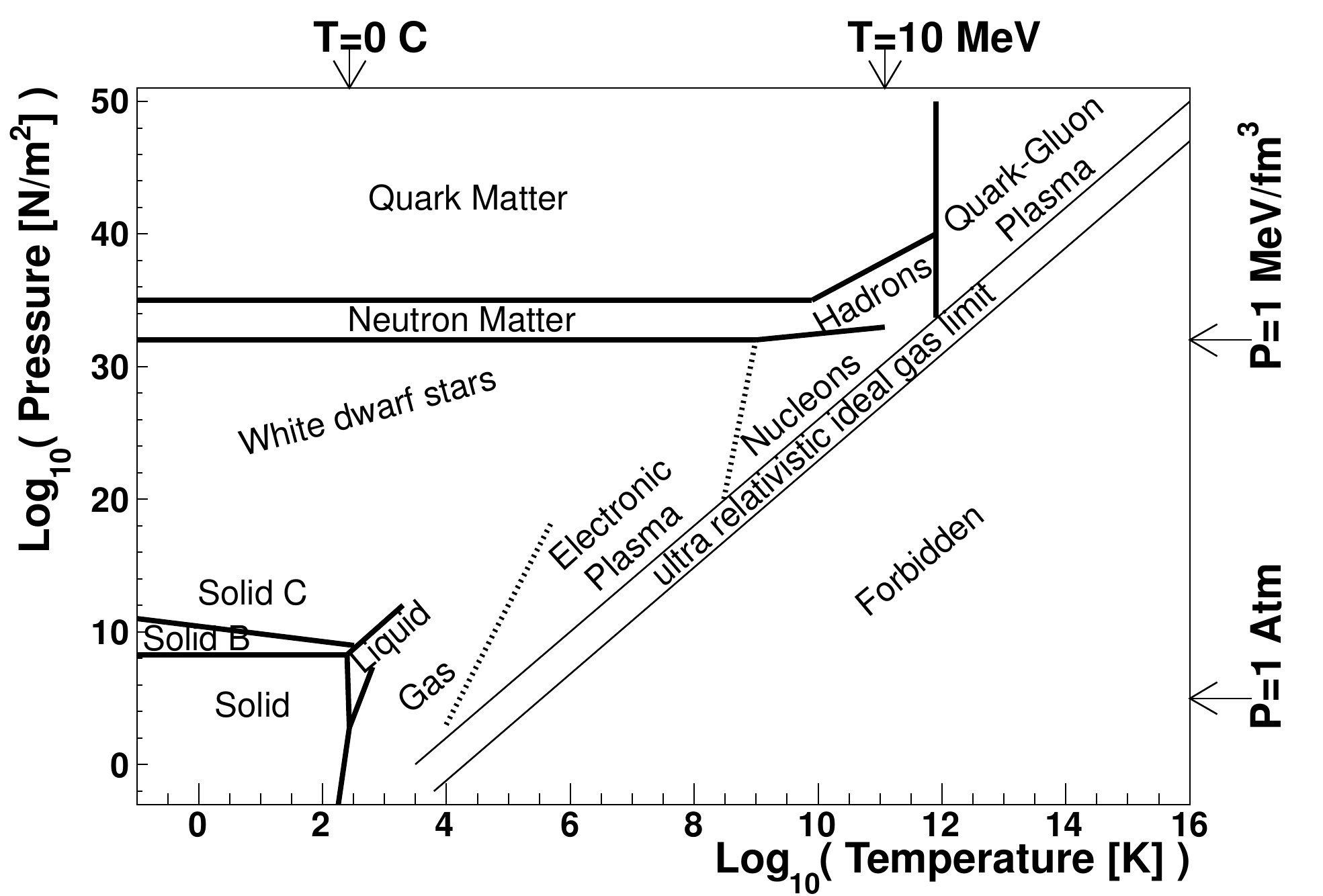}}}}}
\caption{ \label{Intr_DiagPhas}
\emph{Phase diagram of matter in the pressure versus temperature plane for a non zero baryonic potential \cite{Mart06}.}}
\end{figure}

In Fig. \ref{Intr_DiagPhas}, the lay-out of the phase diagram of matter is presented.  We clearly distinguish two regions, one for temperatures below $10^9$ K and pressures below 10$^{30}$ Pa, where the electromagnetic interaction between atoms (or ions) provides the degrees of freedom of matter, and a second region, for temperatures above  $10^9$ K  and/or pressures above $10^{32}$ Pa, where the strong interaction between nucleons, hadrons or quarks dominates and provides the right degrees of freedom. 

 \section{The Quark Gluon Plasma}

\subsection{Limiting temperature of matter}

\begin{figure}
{\centering 
\resizebox*{0.48\columnwidth}{!}{\includegraphics{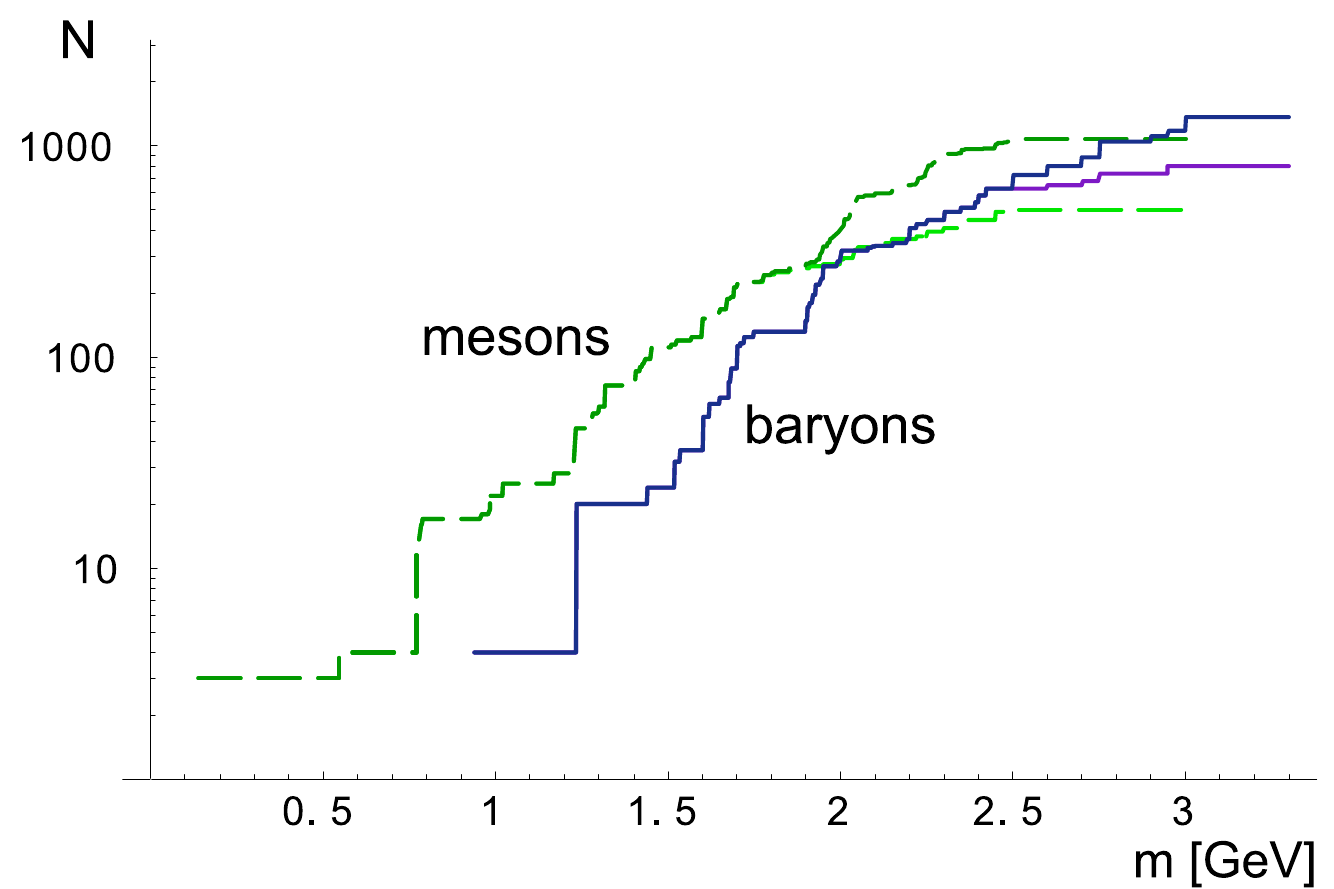}}
\resizebox*{0.48\columnwidth}{!}{\includegraphics{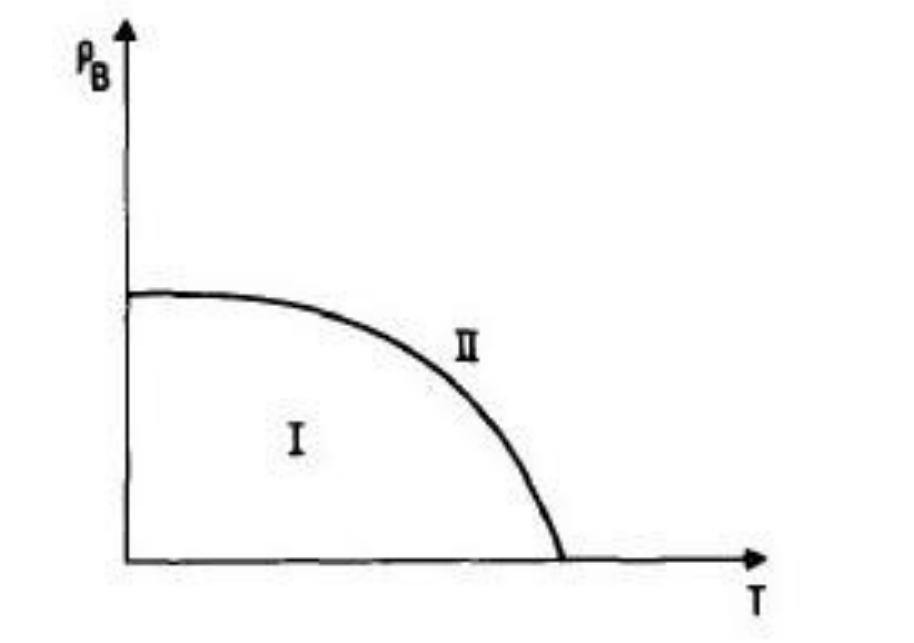}}
\par}
\caption{\label{Tran_1erDiagramme} 
\emph{Left: number of hadron species as a function of their mass \protect\cite{Bron04}.
Right: first phase diagram of hadronic matter \protect\cite{Cabi75}. 
$\rho_B$ is the baryonic density, $T$ the temperature, (I) is the confined phase and  (II) is the deconfined phase.
}
} 
\end{figure}

Curiously, the first prediction of a critical behaviour of hadronic matter at high temperature was obtained before the formulation of QCD and the discovery of partons  \cite{Hage65, Hage84}.
At mid 60's, Hagedorn was interested in properties of the hadron gas. He predicted in a phenomenological and original manner, that should exist a critical behaviour of the hadron gas at high temperature \cite{Hage65}.  Hagedorn interpreted this criticality  as the existence of a maximum temperature of matter, that was called Hagedorn's temperature $T_H$.
In order to study the hadron gas, one has to consider all the \emph{zoology} of hadron particles. Today, more than 2000 hadron species have been discovered  (see Fig. \ref{Tran_1erDiagramme} left). Hagedorn studied the number of hadron species as a function of their mass. He observed an exponential dependence and the following function was used to describe the experimental data:

\begin{equation}
\rho(m) = \frac{A}{m^2+[500 {\rm MeV}^2]}  \exp{ {(m/T_H)} }
\end{equation}
where $\rho(m)$ is the density of hadron species per mass unit and $T_H$ is a parameter. From the experimental data, one obtains that the parameter $T_H$ is close to the mass of the pion, $\sim 180$ MeV, when all the known baryon and meson resonances are considered \cite{Bron04}.
It turns out that such a dependence of the density $\rho(m)$ will induce divergences of the partition function that describes the statistical properties of a hadron gas,  if the temperature of matter reaches values above the Hagedorn parameter $T_H$. 
In consequence, $T_H$ was interpreted as a limiting temperature of matter. Somehow, any additional energy supplied to the system at the Hagerdorn temperature, would be used to create new hadron species.

We know that hadrons are not point-like particles, and their typical size is around 1 fm sphere radius. Indeed, when one gets closer to the $T_H$ temperature, the hadron density increases (remember that the energy density of an ideal ultra-relativistic gas increase as $T^4$) and values about 1 hadron per fm$^3$ are reached. Under these conditions, hadrons overlap with each other and considering hadrons as point-like particles (which means that their size is small with respect to its mean free path) becomes a wrong hypothesis and invalidate Hagedorn conclusions. Therefore one has to understand first the internal structure of hadrons, since it is going to provide the new degree of freedom of the system when T$\geq T_H$. Only QCD was able to answer this question several years later.

\subsection{Deconfined state of matter}
After the discovery of the asymptotic freedom \cite{Poli73,Gros73},  the existence of a deconfined state of quarks and gluons was predicted at high temperature and/or high pressures \cite{Coll75, Cabi75}.  A first pioneer phase diagram of hadronic matter was imagined (see Fig. \ref{Tran_1erDiagramme} right from reference \cite{Cabi75} ). At sufficiently high temperatures, quarks and gluons interact weakly  and the system will behave as an ideal ultra-relativistic gas. The degrees of freedom will be then determined by the flavor numbers, spin states, color and charge states of the quarks and gluons.
The deconfined state was called later quark gluon plasma \cite{Shu78}. The word  \emph{plasma} is used to describe the state of matter when ions and electrons are dissociated in atoms. There is then an analogy when colourless particle dissociate to create deconfined matter. One open question after the discovery of the asymptotic freedom, concerned the properties of the transition from the hadron gas to the QGP: does it take place smoothly or via a phase transition and exhibiting critical behaviours? As a matter of fact, the transition from gas to electronic plasma takes place smoothly in the temperature range 10000 to 50000 K \cite{Stoc99} and no critical behaviour is therefore observed.
The question whether the QGP phase transition exists, is, of course, a very deep question and the intrinsic symmetries of the QCD could give us the answer. Indeed the chiral symmetry of the massless quark QCD Lagrangian is spontaneously broken at low temperature and this symmetry should be restored at high temperatures. A symmetry restoration  represents a valid condition to predict the existence of a QCD phase transition. It remained however an open question if the chiral symmetry transition and the deconfinement transition are or not the same one. Only lattice calculations have been able to provide an answer to this question as we will see later.

 \subsection{The spontaneous break-up of chiral symmetry in QCD}
A simplified Lagrangian of 3 quark flavours $f$ (\emph{u, d ,s}) can be written as \cite{Scha05}:
\begin{equation}
\mathcal{L} =   \sum_f^{N_f} \bar{\psi}_f ( iD\hspace*{-0.27cm}/\, - m_f) \psi_f - \frac{1}{4} G_{\mu\nu}^a G_{\mu\nu}^a,
\end{equation}
where $N_f=3$ and the coupling gluon field  tensor is defined as:
\begin{equation}
 G_{\mu\nu}^a = \partial_\mu A_\nu^a - \partial_\nu A_\mu^a
  + gf^{abc} A_\mu^b A_\nu^c,
\end{equation}
and the covariant derivate of the quark field as:
\begin{equation}
 iD\hspace*{-0.27cm}/\,\psi = \gamma^\mu \left(
 i\partial_\mu + g A_\mu^a \frac{\lambda^a}{2}\right) \psi.
\end{equation}
Under these conditions, the previous Lagrangian exhibits a flavour symmetry since the quark interaction does not depend on the quark flavour. This is indeed always the case if the masses of the quarks are identical. The direct consequence of this is the symmetry under isospin transformations, that it is observed in the hadron properties. In addition, for massless quarks, the QCD Lagrangian exhibits the chiral symmetry\footnote{Chiral from \emph{hand} in Greek.}. The quark fields can be decomposed in left-hand and right-hand quarks fields  \cite{Halz84}:
\begin{equation}
\psi_{L,R} = \frac{1}{2} (1\pm \gamma_5) \psi .
\end{equation}
As a consequence, the QCD Lagrangian is invariant under helicity and flavour transformations. This symmetry is represented as the $SU(3)_L\times SU(3)_R$  symmetry of QCD. One of the consequences of this symmetry is that the associated parameter, called \emph{condensate} $\langle q \bar{q} \rangle$ should be zero.

Nevertheless, the \emph{condensate}  $\langle q \bar{q} \rangle$ is not zero and the existence of the pion is a clear confirmation of this statement \cite{Knec98}. This is what it is called the \emph{spontaneous}  breaking of the $SU(3)_L\times SU(3)_R$ chiral symmetry of QCD. The word \emph{spontaneous} reminds us that the symmetry is respected by the QCD Lagrangian but broken by their states at low energies. At high energies the symmetry should be restored. 

\begin{figure}
{\centering 
\resizebox*{0.55\columnwidth}{!}{\includegraphics{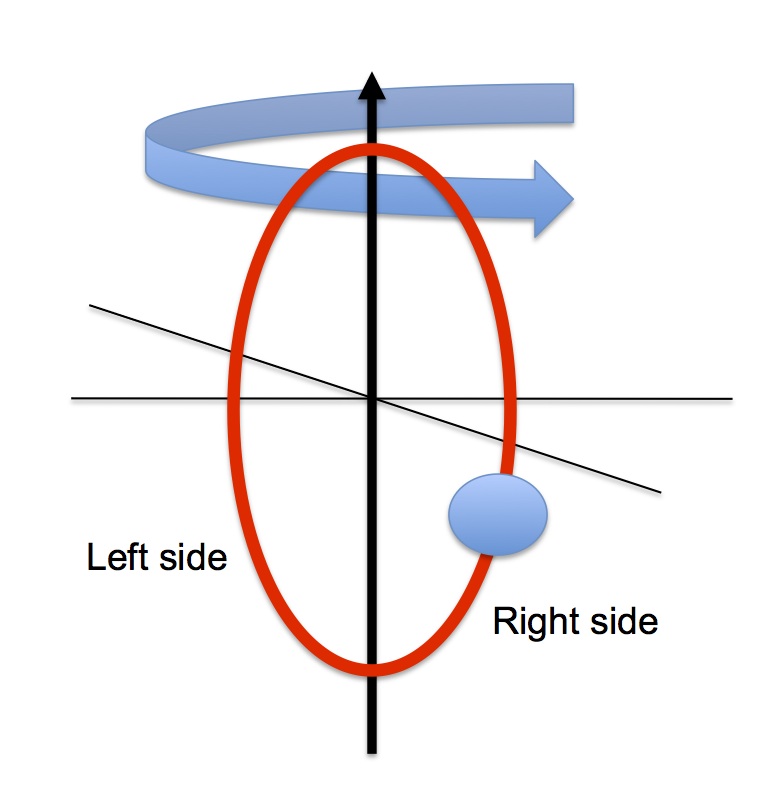}}
\par}
\caption{\label{Intr_ClassicalAnalogy}
\emph{Classical analogy of spontaneous symmetry breaking. The ball is holed and can  move freely along the ring, which rotates with an angular speed $w$, in the earth gravitational field with respect to its vertical symmetry axis. }}
\end{figure}

The spontaneous breaking of a symmetry is a phenomenon that  is allowed in quantum field theories, where the structure of the vacuum plays a major role. In quantum mechanics, the eigenstates that respect the symmetry of the Hamiltonian, can always be found. In classical mechanics the following analogy of the spontaneous symmetry breaking can be found. Let's assume a ring in the earth gravitational field, that  rotates along its vertical symmetry axis with an angular speed $\omega$. There is a small solid ball with a hole in a manner that can move freely along the ring (see  Fig. \ref{Intr_ClassicalAnalogy}).  In this example, the system exhibits a left-right symmetry which is spontaneously broken by the small ball at low internal energy. Indeed, due to the centrifugal force, the small ball has to choose the left or the right side of the ring as its equilibrium position. If some internal energy is given to the ball, it will start to oscillate around its equilibrium position. The amplitude of the oscillation will increase with the internal energy of the ball. Above a certain energy threshold, the ball will have enough internal energy to reach the other side of the ring and it will then move in both sides. When this occurs, one can say that the left-right symmetry of the system has been restored.

The spontaneous breaking of the chiral symmetry is one of the predictions of QCD  \cite{Knec98}, and, in this way, QCD is able to predict the existence of the Goldstone bosons: the pions, kaons and eta mesons and to explain their small interaction cross-sections. As in the classical analogy, the chiral symmetry of the QCD is restored at high energies (or high temperatures) and  remember that a  restoration of the symmetry represents a sufficient condition for the existence of a QCD phase transition.  An analogy with the ferromagnetic phase transition can be made (see table \ref{tab:ana}) \cite{Scha05}.
In fact, the ferromagnetic phase transition can be associated to the spontaneous breaking of the isotropy symmetry. At high energy the ferromagnetic system is invariant under rotation transformation, since there is not any privileged direction of the space. Nevertheless at low temperatures, the thermal agitation cannot avoid that the microscopic magnetic moments of the elementary constituents align, causing a macroscopic magnetisation of the system. Therefore the isotropy symmetry is spontaneously broken  at low temperatures, and this is a sufficient condition to predict that there is a phase transition during the generation of the macroscopic magnetisation of the system. In the ferromagnetic case, the magnetisation  $\vec{M}$ is the order parameter of the transition, which is the equivalent of the quark \emph{condensate} $\langle q \bar{q} \rangle$  in the chiral transition in QCD.
The non-zero $\vec{M}$, allows for the existence of spin waves, and the Goldstone bosons  (pions, kaons and eta's) are their analogous. Finally, isotropy symmetry can be explicitly broken via an external magnetic field. The equivalent of the non-zero external magnetic field would be the non-zero masses of the quarks, which explicitly breaks the chiral symmetry of the QCD Lagrangian.

\begin{table}
\begin{center}
\begin{tabular}{|c|c|c|} \hline \hline
Transition & Chiral & Ferromagnetic\\ \hline
Spontaneous breaking& $SU(3)_L\times SU(3)_R$  & Isotropy O(4) \\ 
Order parameter & Condensate $<q\bar{q}>$  &  Magnetisation $\vec{M}$ \\
States & Goldstone bosons $\pi$, $K$, ... & Spin waves \\
Explicit breaking & Quark masses $m_q\neq0$ & External magnetic field \\ \hline
\end{tabular}
\end{center}
\caption{Analogy between the chiral  and the ferromagnetic phase transition  \protect\cite{Scha05}.}
\label{tab:ana}
\end{table}

We have seen that a spontaneous breaking of the chiral symmetry explains why there should be a phase transition of the hadronic matter. We can now wonder if such a transition is associated to the process of deconfinement of quarks and gluons leading to the formation of a quark gluon plasma. One could imagine that there are, indeed, two different phase transitions, a chiral transition and a deconfinement one that occur at different critical temperatures. In the next section, lattice QCD calculations will be presented since this is the only way to answer this question today.

It should be noted that we have assumed a QCD Lagrangian with massless $u$, $d$ and $s$ quarks.
This is indeed a good approximation, since the masses of the, so called light quarks, are small compared to $\Lambda_{QCD}$ but they are not zero:
$m_u=2.3 \pm 0.5$ MeV, $m_d=4.8^{+0.7}_{-0.3}$ MeV and $m_s=95\pm5$ MeV \cite{PDG12}.
In this respect the chiral symmetry is indeed explicitly broken by the QCD Lagrangian. Above we have assumed that if the masses are small compared to $\Lambda_{QCD}$ this chiral symmetry should remain a good symmetry of QCD. However this may be a wrong assumption, in particular for the strange quark. Indeed, it is an open question what would be the masses value thresholds causing the damp out of the criticalness of the chiral transition. Above such mass thresholds, the chiral transition would become a cross-over and no critical behaviour would be observed in the transition. Once more, the lattice QCD calculations will be a unique method to study this question.

Finally, there is a new symmetry of the QCD Lagrangian in the limit of quarks masses $m_q \rightarrow \infty$. The order parameter of this symmetry is called the Polyakov line $\langle P \rangle$ which is directly associated to the process of deconfinement if $\langle P \rangle$=0 \cite{Scha05}.

\subsection{Some results from lattice QCD calculations at finite temperature} 
\label{lattice}

Today, lattice QCD calculation is a unique method to test QCD in the non-perturbative domain.
In the last decades, many progresses have been achieved on the algorithms and on the computing performances.
Lattice QCD allows for non-perturbative calculations with high reliability.

\begin{figure}
{\centering 
\resizebox*{0.90\columnwidth}{!}{\includegraphics{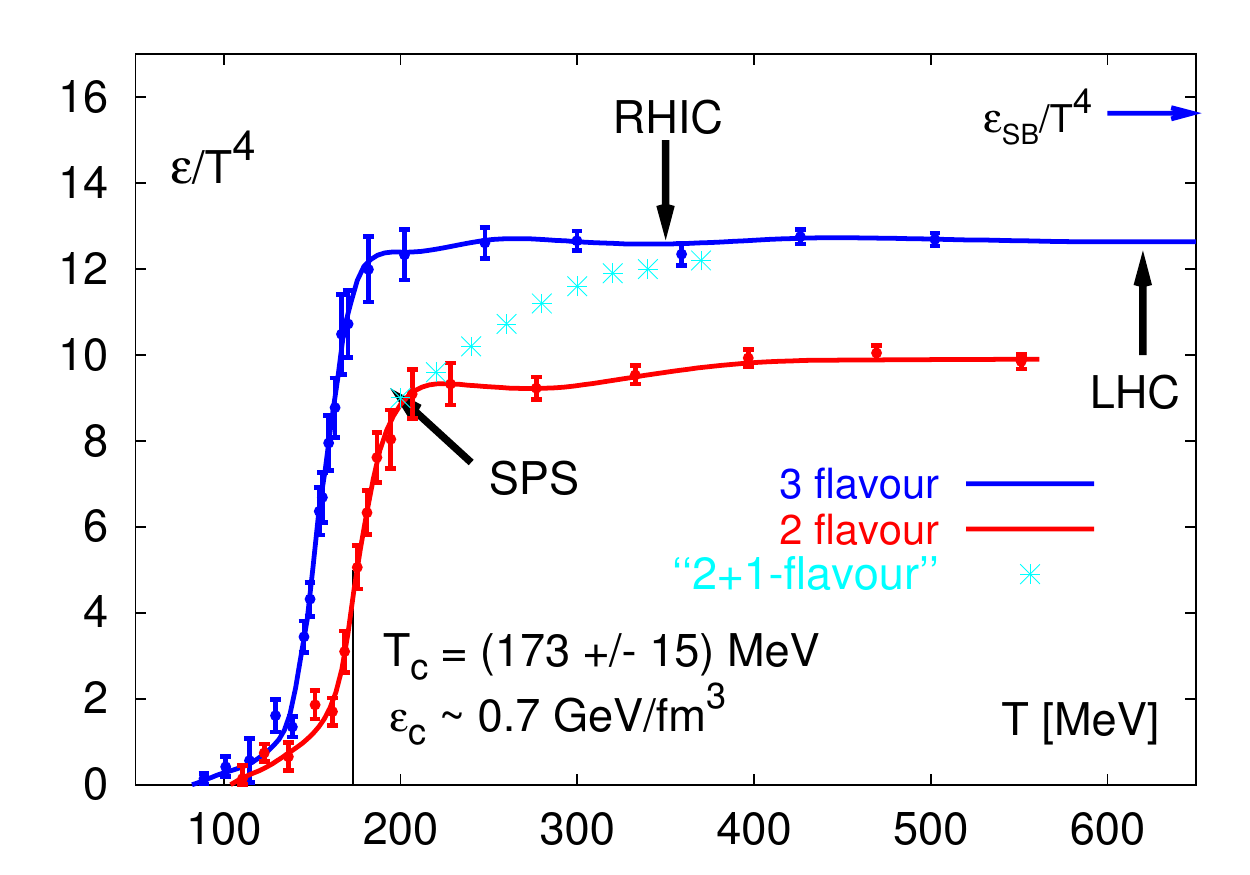}}
\par}
\caption{\label{Tran_QCDReseau}
\emph{Dependence of the energy density as a function of the temperature of the hadronic matter at null baryonic potential given by lattice QCD calculations at finite temperature. The calculations are performed for two massless quarks, three massless quarks and two massless quark and one ($s$)  with its real mass. A transition is observed at a temperature of about 173 MeV and energy density of 0.7 GeV/fm$^3$.  For the calculations with a  real $s$ mass, the transition is faded away  \cite{Kars01b}}} 
\end{figure}

In particular, lattice QCD should allow to study the properties of the Universe between few ns and few $\mu$s  after the Big Bang (temperatures around 100-1000 MeV) and to study hadronic matter in the core of the neutron stars \cite{Petre12}. For massless quarks, these calculations show a transition at baryonic potential $\mu_B=0$, as expected from the spontaneous breaking of the chiral symmetry in QCD. The critical temperature would be  $T=173\pm15$~MeV and the critical energy density $\epsilon =0.7\pm0.3$~GeV/fm$^3$ \cite{Kars01b} (see Fig.  \ref{Tran_QCDReseau}). It is also observed that above the critical temperature, the energy density is indeed proportional to $T^4$, as  for an ideal ultra-relativistic gas, but the proportionality factor (Stefan-Boltzmann constant) is about 20\% smaller than the expected value for an ideal gas of  gluons and massless $u$, $d$ and $s$ quarks. Perturbative calculations at higher temperatures are able to explain the evolution of this factor for $T\geq 2T_c$ \cite{Blai99}. 

\begin{figure}
{\centering 
\resizebox*{0.90\columnwidth}{!}{\includegraphics{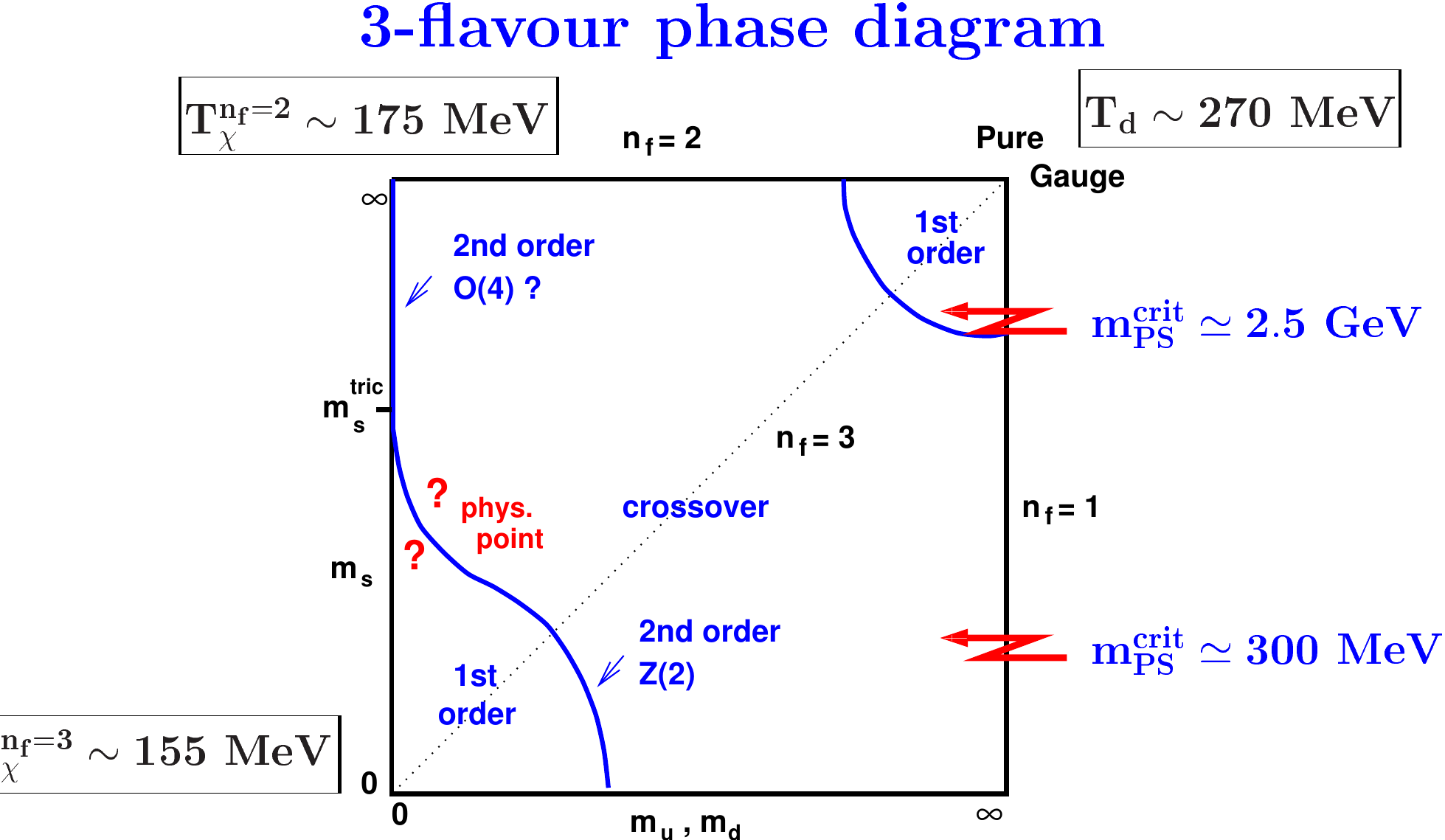}}
\par}
\caption{\label{Tran_QCDReseauMasses}
\emph{Lattice QCD calculations  of the criticalness of the hadronic matter phase transition for 3 quark flavours, $\mu_B=0$ and assuming the mass of the $u$ and $d$ quarks are identical and a strange quark mass, $m_s$ \protect\cite{Kars01, Kars01b}.}}
\end{figure}

The lattice QCD calculations show that for massive quarks, the phase transition could fade away, it would become a cross-over and no criticalness would be observed.  The criticalness of the transition has been studied as a function of the quark masses (see Fig. \ref{Tran_QCDReseauMasses}). In the calculations presented here, the $u$ and $d$ masses are considered to be identical and $\mu_B=0$. It is observed that for both low and large masses, a 1st order phase transition is predicted. The cross-over transition occurs for intermediate quark masses. A 2nd order phase transition occurs in the border line between 1st order and cross-over areas. Today there is some consensus to believe that for the physical quark masses and $\mu_B$=0 there is not a phase transition but a cross-over \cite{Kars01b, Kars01}\footnote{Note that more recent references on this subject exist and they are not referenced in this lecture.}. 

The QCD lattice calculations with physical quark masses, have determined critical temperatures between 150-200 MeV. There has been some confusion about the exact critical temperature of the transition in the last years. The outcome was that the evaluation of the transition temperature, which is not a well defined parameter for a cross-over transition, would depend on the method used for its determination. Calculations based on chiral order parameter show a cross-over transition for T$\sim$155 MeV. On the other hand,  the behaviour of the Polyakov loop suggests that colour screening sets in at temperatures that are higher than the chiral transition temperature \cite{Petre12}.

\begin{figure}
{\centering 
\resizebox*{0.45\columnwidth}{!}{\includegraphics{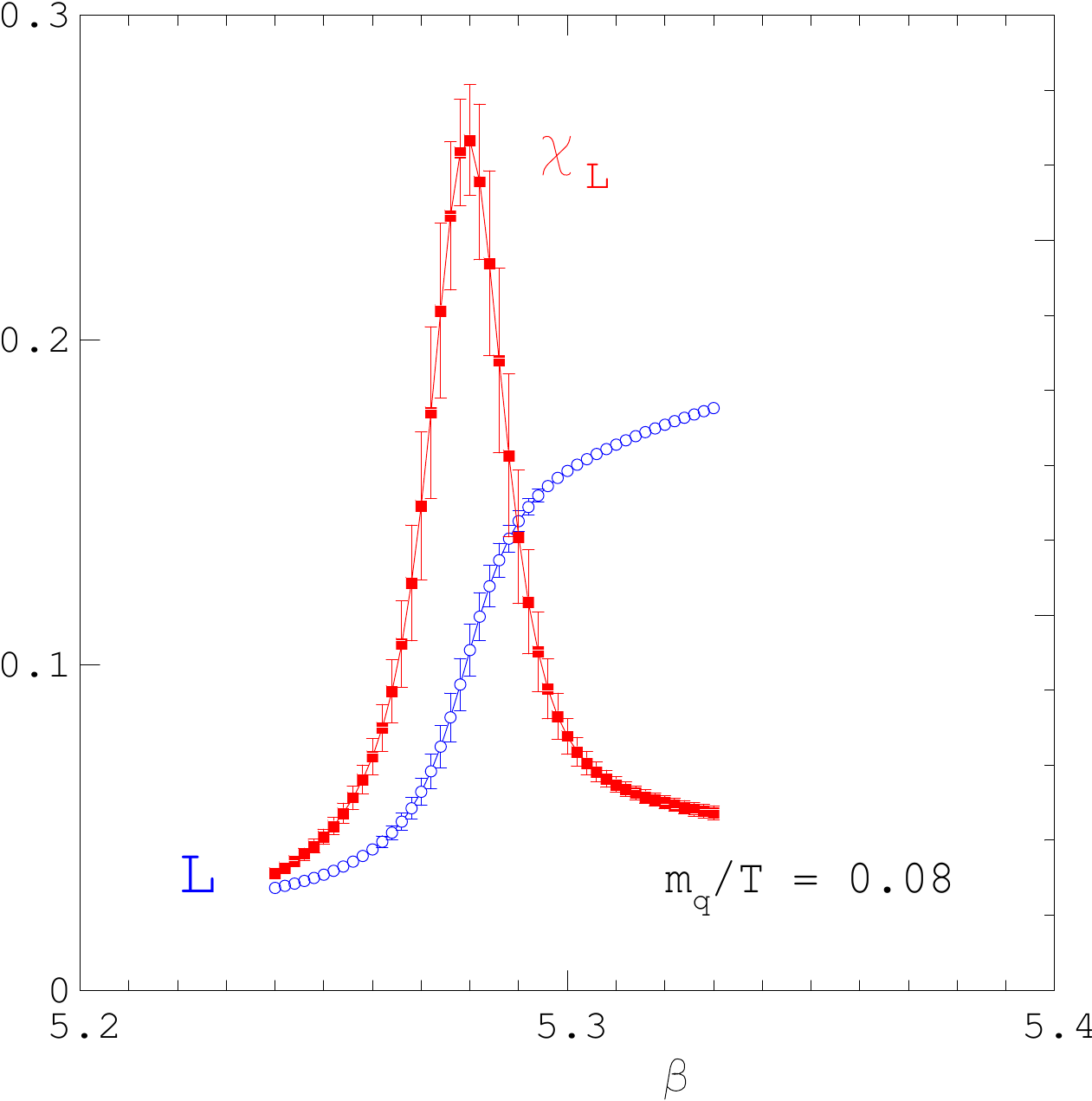}}
\resizebox*{0.45\columnwidth}{!}{\includegraphics{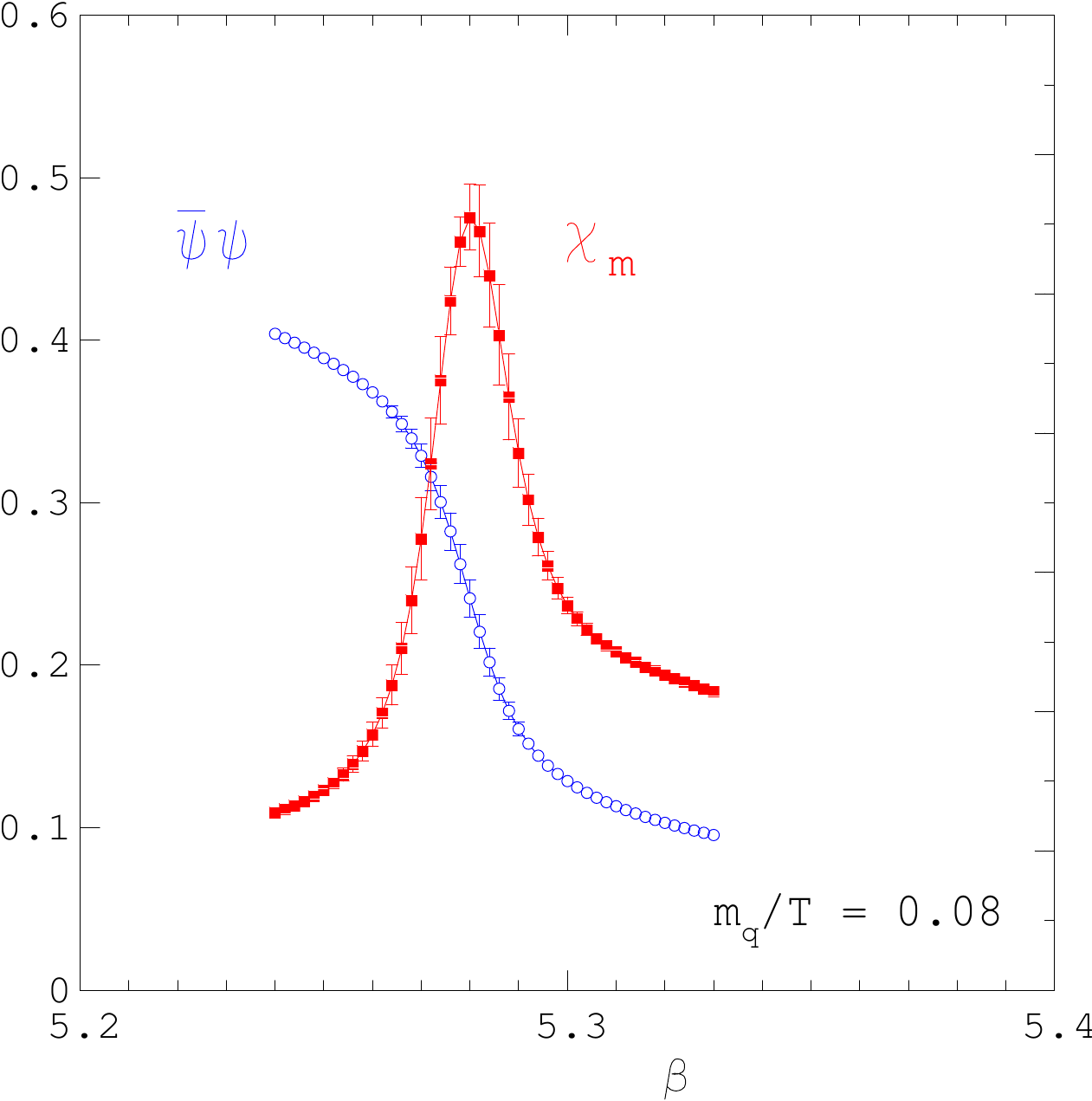}}
\par}
\caption{\label{Tran_QCDReseauChirale}
\emph{Critical behaviour for massless quarks and $\mu_B=0$ of the order parameters of the deconfinement (left plot)  and of the chiral (right plot) transitions as predicted by lattice QCD calculations. The order parameters are the Polyakov susceptibility ($\chi_L$) and the chiral susceptibility ($\chi_m$) \protect\cite{Kars01b}. Both transitions would indeed be the same one or would take place at the same critical temperature.}
}
\end{figure}

Finally, lattice QCD calculations have studied the order parameters of the chiral and deconfinement transitions  (see Fig. \ref{Tran_QCDReseauChirale}) showing that, a priori, both transitions occur at the same critical temperature. Therefore, this suggests that both transitions would be indeed the same transition. However, the interplay between chiral and deconfinement aspects of the transition appears to be more complicated than earlier lattice studies suggested. It seems that there is no transition temperature that can be associated with the deconfining aspects of the transition for physical values of the light quark masses \cite{Petre12}.

\begin{figure}
{\centering 
\resizebox*{0.90\columnwidth}{!}{\includegraphics{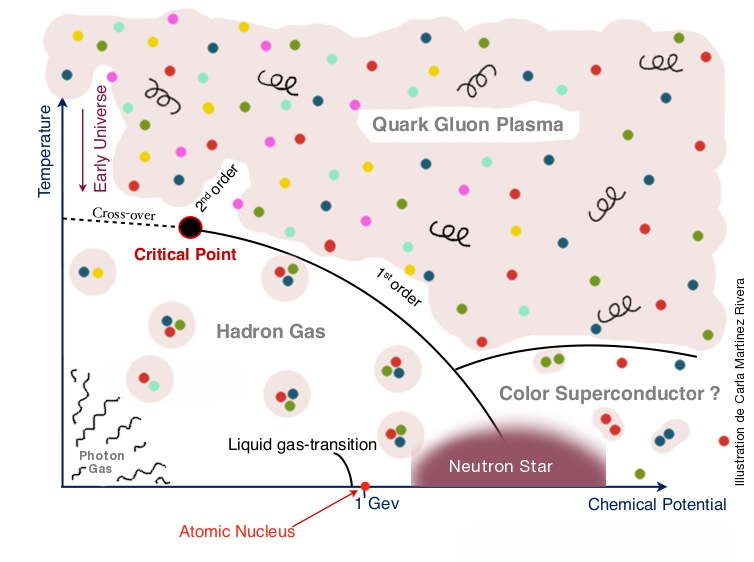}}
\par}
\caption{\label{Tran_Schema}
\emph{Lay-out of the hadronic matter phase diagram as it is today conceived.}
}
\end{figure}

In the last decade, a lot of effort has been done to perform calculations at $\mu_B\ne0$. These calculations show that there would be a critical point at $\mu_B\sim0.75M_N$ ($M_N$ is the nucleon mass) where the cross-over becomes a 2nd order phase transition, and beyond it, the transition becomes a 1st order phase transition between the gas of hadrons and the quark gluon plasma \cite{Fodo02}. 
In addition, other calculations have predicted a transition to a colour superconductor matter at high values of $\mu_B$ (see the lay-out of the hadronic matter phase diagram in Fig.  \ref{Tran_Schema}).

\subsection{Properties of a QGP in the ultra-relativistic limit}

At temperatures $\Lambda_{QCD} \le T  \le$ charm mass, $m_c$ and assuming that the strong interaction strength becomes very small, the QGP would behave as an ideal gas. Strictly speaking this gas will be constituted  by all the elementary particles ($m\lesssim T$): leptons (electrons and muons), bosons (photons and gluons) and light quarks $u$, $d$ and $s$ \footnote{We assume a temperature lower than the mass of the charm quark and in this manner one could neglect the thermal production of heavy quarks, tau leptons and weak bosons} and their corresponding antiparticles. This gas will have similar properties as the black body radiation and the equation of state is given by  $\epsilon = 3p$, where $\epsilon$ is the energy density and $p$ is the pressure. The energy density will depend on the temperature following the Stefan-Boltzmann law $\propto T^4$. The Stefan-Boltzmann law for bosons is \cite{Land67,Grei95}:
\begin{equation}
\epsilon_b = 3p = g \frac{\pi^2}{(\hbar c)^3}\frac{(k_B T)^4}{30}
\end{equation}
where $g$ is the number of degrees of freedom due to spin, flavour, and colour charge of the considered particle.
In consequence, $\epsilon/T^4$ or $p/T^4$  will be constant for such a matter.
If one only considers photons (black body radiation) we obtain the Stefan-Boltzmann constant ($g$=2 for the two possible spins of the photon):
\begin{equation}
\sigma  = \frac{\pi^2k_B^4}{60\hbar^3c^2}=5.670 \cdot 10^{-8}~{\rm W m}^{-2} {\rm K}^{-4}.
\end{equation}
In natural units (temperature in MeV), the equation of a photon gas is
\begin{equation}
\epsilon_\gamma = A \times T^4 ~ [{\rm MeV}^4]
\end{equation}
with $A\sim0.65$. 
The Stefan-Boltzmann law for fermions is similar to that for bosons \cite{Land67,Grei95}:
\begin{equation}
\epsilon_f = 3p = g \frac{7\pi^2}{(\hbar c)^3}\frac{(k_B T)^4}{240}.
\end{equation}

The total energy density of this matter will be:
$\epsilon = \epsilon_\gamma + \epsilon_l + \epsilon_g + \epsilon_q$;
with $g=8$ for leptons (2 for spin, 2 for flavors and 2 particle-antiparticle), $g=16$ for gluons (2 helicity states and  8 colour charges) and $g=36$ for quarks (2 for spin, 3 colours, 3 flavours, and 2 particle-antiparticle):
\begin{equation}
\epsilon = (A_\gamma +A_l+A_g+A_q)\times T^4 ~ [{\rm MeV}^4]
\end{equation}
with $A_\gamma$=0.65, $A_l$ = 2.30, $A_g$=5.26 and $A_q$=10.36.

For a small size plasma (radius below $10^{-10}$ m) or short lifetime, electromagnetic particles like photons and leptons could not reach thermalisation. They will be radiated by the thermalised medium but they will not be in equilibrium with the medium. Ignoring them, one gets $\epsilon = 15.62\times T^4 ~ [{\rm MeV}^4]$ for 3 flavors of massless quarks and 8 gluons (see the value of $\epsilon_{SB}/T^4$  in Fig. \ref{Tran_QCDReseau}).

\subsection{Probes of the QGP}

\subsubsection{Thermal radiation}
\label{thermalradiation}

Thermal radiation from a QGP will allow to study several properties of the QGP like its temperature $T$. On the surface of the QGP volume, photons\footnote{but also the other fundamental particles.} will escape. This is the thermal radiation.

\begin{figure}
{\centering 
\resizebox*{0.90\columnwidth}{!}{\includegraphics{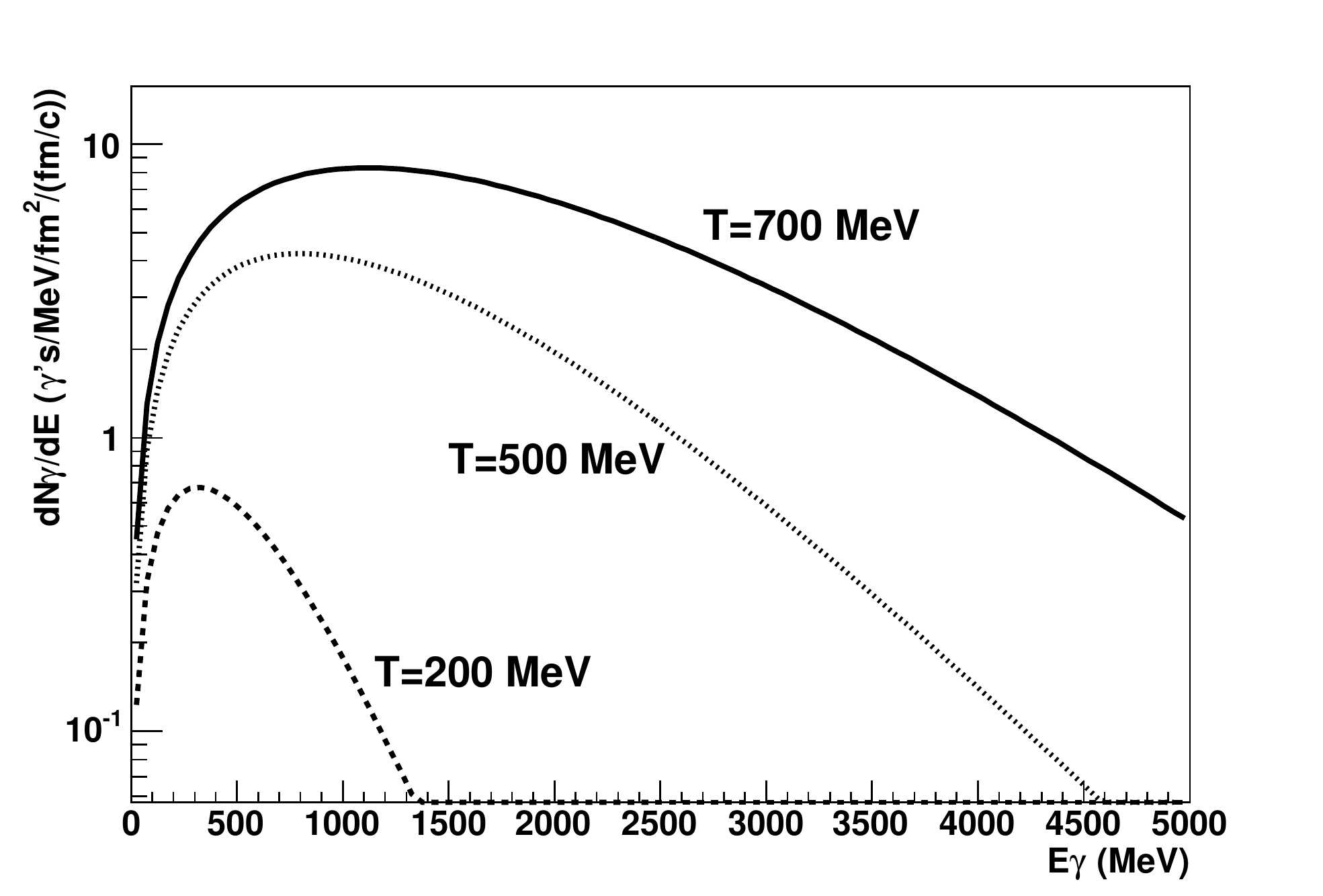}}
\par}
\caption{\label{Tran_CorpNoir}
\emph{Thermal photon production in a  QGP of 7 fm radius and during 10 fm/c, as expected from a black-body radiation.}}
\end{figure}

As we have estimated for an ideal QGP, the partial pressure of these photons on the QGP surface, will be given by the expression $p = \epsilon/3 =0.22T^4$ and their energy distribution by the Planck law. In consequence the differential partial pressure  $dp/dE_\gamma$  in natural units will be :
\begin{equation}
\frac{dp(E_\gamma, T)}{dE_\gamma} = 0.034 \frac{E_\gamma^3}{\exp{(E_\gamma/T)} -1}~[{\rm MeV}^3]
\end{equation}
where $E_\gamma$ is the photon energy in MeV.
Considering massless particle and a QGP of a radius 7 fm and 10 fm/c lifetime, the thermal radiation spectrum is presented in Fig.  \ref{Tran_CorpNoir} for temperatures of 200 MeV, 500 MeV and 700 MeV. The corresponding photon yields for energies above 1 GeV are 52, 4600 and 16000, respectively.

Obviously, the numerical example presented here is unrealistic since a 7 fm radius QGP will be transparent to photons. Under these conditions, the electromagnetic radiation of a thermalised QGP is not in thermal equilibrium with the medium which is producing it. Once a photon is produced, it will escape from the QGP, therefore the emission is from the volume and not from the surface as in the black-body radiation.  The calculation of the thermal photon radiation from a QGP is complicated  \cite{Geli03, Yell03b}. At first order, one could expect a reduction of the total number of photons emitted following the ratio of the strength of the strong and the electromagnetic forces $\alpha_{QED}/\alpha_{QCD}$. Only for large size QGP, with a radius above $\sim$0.1 \r{A}, the black-body radiation model would become valid.

\subsubsection{Screening of the colour potential between heavy quarks in the QGP}

As we have already mentioned, the transition to the QGP only concerns the light quarks $u$, $d$ and $s$, for which the chiral symmetry is a good approximation. Since heavy quarks explicitly break the chiral symmetry, they are not directly concerned by the transition to QGP. In other words, the bound states of heavy quarks (quarkonia) are not necessarily melt in a QGP and they could exist as bound states. For this reason, these bound states become very interesting probes for measuring the temperature of the QGP \cite{Satz86}. 

Let us see qualitatively which are the properties of a quarkonium embedded in a QGP. 

Quarkonia are bound states between two heavy quarks $Q\bar{Q}$:  $c\bar{c}$ for the family $\eta_c$, J/$\psi$, $\psi(2S)$, $\chi_c$ ... and the states $b\bar{b}$ for the family $\Upsilon$'s and $\chi_b$. The bound state $t\bar{t}$ has not been experimentally observed and  it will surely not exist due to the short lifetime of the top quark. Finally, one should note that the hadronic states $c\bar{b}$ ($\bar{c}b$)  will have similar properties as quarkonium, although their decay should be similar to that of a $B$ hadron. In vacuum, the quarkonium spectrum can be described via non relativistic models based on a potential interaction like:
\begin{equation}
V(r) = \sigma r - \frac{\alpha}{r}
\end{equation}
where  $\sigma$ represents the string tension  $Q\bar{Q}$ and $\alpha$ is a Coulombian-like constant \cite{Satz86}.
For simplicity, let us assume that the potential is only Coulombian, so $\sigma=0$.

If the $Q\bar{Q}$  state is embedded in a QGP at a temperature $T$, the interaction potential between the heavy quarks will be affected by the presence of the free colour charges in the QGP. This is the screening of the potential. This phenomenon is well known in electromagnetic plasma.
In the plasma, the Coulombian potential has to be replaced by a potential with a screening constant:
\begin{equation}
V(r) = - \frac{\alpha}{r} \times e^{(-r/\lambda_D)}
\end{equation}
where $\lambda_D$ is the Debye length.
Let us assume that the average distance between the heavy quarks in a 1S quarkonium state (J/$\psi$ or $\Upsilon$(1S)) can be estimated by the Bohr radius expression:
\begin{equation}
r_B = \frac{1}{\alpha m_Q}.
\end{equation}
As a numerical example, one can consider for the  J/$\psi$  $m_c$=1250~GeV and $\alpha(m_c)=0.36$ \cite{PDG12}, so $r_B=0.44$~fm.
For the $\Upsilon(1S)$,   $m_c$=4200~GeV and $\alpha(m_b)=0.22$ \cite{PDG12}, so $r_B=0.22$~fm.

If $r_B \ll \lambda_D$, the potential between the heavy quarks can be considered as a Coulombian potential and the bound state exhibits the same properties in the QGP as in the vacuum. However, if $r_B \geq \lambda_D$, the quarkonium properties will be modified by the medium, and it could happen that the quarkonium becomes an unstable state and therefore would melt.
For electromagnetic plasmas, the Debye length depend on the temperature of the plasma and the charge density $\rho$ \cite{Stoc99}:
\begin{equation}
\lambda_D = \sqrt{\frac{T}{8\pi\alpha \rho}}
\end{equation}
Assuming that the previous expression is also valid for the QGP\footnote{This assumption is not justified, but the conclusions that will be obtained are still valid.} and an ideal ultra-relativistic gas $\rho \propto T^3$, one obtains :
\begin{equation}
\lambda_D \sim \frac{1}{\sqrt{8\pi\alpha} T}.
\end{equation}
And therefore, the quarkonium could be melt for temperature above $T_d$:
\begin{equation}
T_d \sim \frac{1}{\sqrt{8\pi\alpha(T)} r_B}.
\end{equation}
For $\alpha(T)\sim 0.2$, one obtains that  $T_d\sim$~200~MeV (1.3$T_c$) for the J/$\psi$ and 
$T_d\sim$~400~MeV (2.6$T_c$) for the $\Upsilon$(1$S$).  Assuming that for 2S states the $r_B$ would be twice larger, one would conclude that the dissociation temperature for $\Psi$' is $\lesssim T_c$ and for $\Upsilon$(2$S$) similar to that of J/$\psi$.

Of course, these calculations are qualitative and obtained without much detail, but they allow to show what is the role of parameters like the mass of the quarkonium and the strength of the interaction. This explains why $\psi$(2S) resonance is easily melt with respect to J/$\psi$ and why $\Upsilon(1S)$ would melt at higher temperatures than that of J/$\psi$. You will find a rigorous calculation of an upper bound of the dissociation temperature of quarkonium in the contribution of A. Mocsy and P. Petreczky in the reference \cite{Abre08}. Their dissociation temperatures are quoted in table \ref{Tab:DissoTemp}.

\begin{table}
\begin{tabular}{|c|cccccc|} \hline
Bound state    & $\chi_c$ & $\psi$'  & J/$\psi$     & $\Upsilon(2S)$ & $\chi_b$   &  $\Upsilon(1S)$  \\ \hline
$T_d$             & $\lesssim T_c$ & $\lesssim T_c$ & $\sim$1.2$T_c$  & $\sim$1.2$T_c$          & $\sim$1.3$T_c$  & $\sim$2.0$T_c$ \\
\hline
\end{tabular}
\caption{Upper bound of dissociation temperatures $T_d$ of quarkonium states in units of the QGP transition temperature $T_c$ obtained by A. Mocsy and P. Petreczky in \cite{Abre08}.}
\label{Tab:DissoTemp}
\end{table}

\subsubsection{Parton - QGP interaction} 
\label{partonQGP}

The QGP could also be studied via its tomography using high energy partons. QCD predicts that high energy partons will lose energy via gluon radiation when crossing the QGP.
The order of magnitude of the parton energy-loss in QGP would be about $\Delta E\sim$1~GeV/fm and it is expected to be proportional to the gluon density.  In addition QCD also predicts that the formation length of the radiated gluon will be larger than the average distance between the gluons in the QGP (interaction centres of the incident high energy parton).  As a consequence several interaction centres will participate in the gluon emission from the parton, and the amplitude from the interaction centres will interfere (this phenomenon is called Landau-Migdal-Pomeranchuck effect) since the radiated gluon will be coherently emitted along all its formation length. For this reason for QGP thicknesses about 1-3 fm, the $\Delta E$ should be proportional to the square of the transversed path length in the QGP \cite{Baie97,Zakh97}:
\begin{equation}
\Delta E \sim \alpha_s \times C_R \times \hat{q}(\rho_g) \times L^2
\end{equation}
where $\alpha_s$ is the strength of the strong interaction, $C_R$ is the colour charge factor $\hat{q}$ is the transport coefficient which depends on the gluon density ($\rho_g$) of the QGP and $L$ is the thickness of the QGP.

The energy lost will depend on the nature of the parton:
\begin{itemize}
\item Gluons will exhibit larger energy-loss per unit of length than that of quarks. A relative factor 9/4 due to the colour charge, is associated to the gluonsstrahlung mechanism from a gluon with respect to that from a quark  \cite{Peig06}.
\item Heavy quarks are expected to lose less energy than light quarks, due to the absence of gluon radiation at forward angles, below $\theta<M/E$, where $M$ is the quark mass and $E$ its energy \cite{Doks01}. This phenomenon, predicted by the QCD, is called dead-cone effect. The dead-cone effect should become measurable for beauty quarks, whereas this effect should remain relatively small for charm quarks. Moreover, elastic collisions with partons in the QGP could also contribute to the energy-loss of heavy quarks in the QGP. Finally the hadronization time scale for heavy quark hadronization increases due to its larger mass and it could occur, namely for the beauty, that hadronization takes place when the heavy quark is still traversing the QGP. 
\end{itemize}

One can wonder if other high-energy elementary particles like photons, electrons, electroweak bosons etc... could also be used to study the QGP. Photons and electrons will only interact electromagnetically and they should lose energy like in ordinary matter via bremsstrahlung emission and the production of electrons and positron pairs. However, the expected energy-loss is relatively small for QGP of a radius of tens of femtometers, about $\sim$1\% of the energy of the particle \cite{Peig06}.  In the case of electroweak bosons, they will decay quickly due to their short lifetime and only their daughter particles will interact with the QGP.

\section{Heavy Ion Collisions and Heavy Ion Accelerators}
The study of hadronic matter in the laboratory is one of the challenges of experimental nuclear physics since the eighties.
Today, the unique experimental method consists in accelerating and colliding two heavy nuclei.
In laboratories like CERN (Geneva, Switzerland), BNL (New York, USA), GSI (Darmstadt, Germany), and GANIL (Caen, France), nuclei are accelerated at energies that range from MeV to TeV beam energies. Depending on the center-of-mass energy of the collision, different domains of the phase diagram of hadronic matter can be studied. Before the collision, the nucleus-nucleus system is out of equilibrium. During the collision, the strong interaction between the constituents may dissipate a fraction of the available center of mass energy into the internal degrees of freedom of the system, and hopefully, a microscopic drop of hot hadronic matter could be created in the laboratory.  The pressure gradient between the drop and the surrounding vacuum would be incredibly high and the drop will suffer a dramatic expansion against the vacuum. The temperature of the system will change during the expansion and a series of ephemeral thermodynamical states will be created. The complexity of this dynamical evolution of the system makes much more difficult the study of the intrinsic properties of the hadronic matter and a rigorous methodological approach has to be undertaken:

\begin{itemize}
\item \textbf{Collision dynamics.} The systematic study of the different colliding systems, center of mass energies and impact parameter will be of vital importance;

\item \textbf{Experimental probes.} This implies that one can detect, identify and measure the kinematic properties of all the particles produced in the nucleus-nucleus collisions. This has not been always possible, and only large scale experiments in colliders like STAR, PHENIX, ATLAS, ALICE or CMS  are able to perform such a complete measurement. This is the only way to measure all the experimental probes, like particle multiplicity, light and strange hadron yields, transverse momentum and rapidity distributions, hadron correlations, azimuthal asymmetries, heavy quarks, quarkonia, direct photons, jets, dijets, electroweak bosons,  photo-jet and electroweak bosons-jet correlations, etc ... 

\item \textbf{Experimental probes in cold nuclear matter}. In addition, one has to study the experimental probes when the microscopic drop of hot hadronic matter is not created, namely, in peripheral collisions and/or induced-proton collisions;

\item \textbf{Global interpretation.} The results obtained have to be interpreted in one single scenario that explains coherently the whole phenomenology of experimental results.
\end{itemize}

In order to create a drop of QGP in the laboratory, energy densities of about 1.0~GeV/fm$^3$ have to be reached. Nucleus-nucleus collisions at relativistic energies have become a unique experimental method. Naively, one could assume that all the available energy in the center-of-mass is dissipated, during the collision, into the internal degrees of freedom of the nucleus-nucleus system. The latter statement is certainly a bad hypothesis since anyone will expect that a non negligible fraction of the available energy in the center-of-mass will not be dissipated to create hot matter. But this hypothesis allows to estimate the beam energy below which, the QGP cannot be formed. 
Under this hypothesis, the energy density of the drop will approximately be given by
\begin{equation}
\label{Coll_Ecm}
\epsilon \approx \frac{\sqrt{2E_b \times m} \times  A}{V}
\end{equation}
where $m$ is the mass of the nucleon, $E_b$ is the beam energy in the reference system where one of the nucleus is at rest,  $A$ the atomic number and $V$ the initial volume of the system. Assuming:
\begin{equation}
\label{Coll_Volu}
V \approx 4/3\pi \times (1.124)^3 \times A~[{\rm fm^3}], 
\end{equation}
we conclude that for beam energies $E_b$ below $\sim$20 GeV per nucleon (that is, a center-of-mass energy below $\sqrt{s_{\rm NN}}$=6 GeV per nucleon pair) the energy available in the center of mass is insufficient to heat a nucleus to energy densities above 1~GeV/fm$^3$. It seems hardly possible that a drop of QGP could be then formed.  Even at $E_b$ $\sim$ 20 GeV per nucleon,  the stopping power of nuclear matter would not be strong enough to stop both nuclei, and in consequence only a fraction of the initial beam energy could be dissipated into the internal degrees of freedom of the system. Therefore $E_b$ noticeably larger than 20 GeV per nucleon would be needed to reach the critical density of the QGP phase transition.

At the beginning of the 80's, the American physicist J.D Bjorken imagined a scenario where the QGP would be efficiently formed. He described what would be the initial energy density and its evolution with time  \cite{Bjor83}. As we will see later, one of the hypothesis of this scenario is only corroborated for $E_b$ larger than  $\sim$250 GeV, that is an available energy in the center of mass larger that 25 GeV per nucleon pair. In addition, the QGP would be formed at baryonic potentials close to zero under this scenario.

Therefore, it remains an open question whether the critical energy density could be reached or not, in the intermediate domain between $E_b$=20-250 GeV ($\sqrt{s_{\rm NN}}$=6-25 GeV).  As I will mention later,  the results from the SPS experimental heavy ion program (1986-2000) at $\sqrt{s_{\rm NN}}$=17-19 GeV, hinted at the existence of a new state of matter in which quarks, instead of being bound up into more complex particles such as protons and neutrons, are liberated to roam freely.

\subsection{The Bjorken scenario of heavy ion collisions at ultra-relativistic energies}
\label{bjorkenscenario}

\begin{figure}
{\centering 
\resizebox*{0.70\columnwidth}{!}{\includegraphics{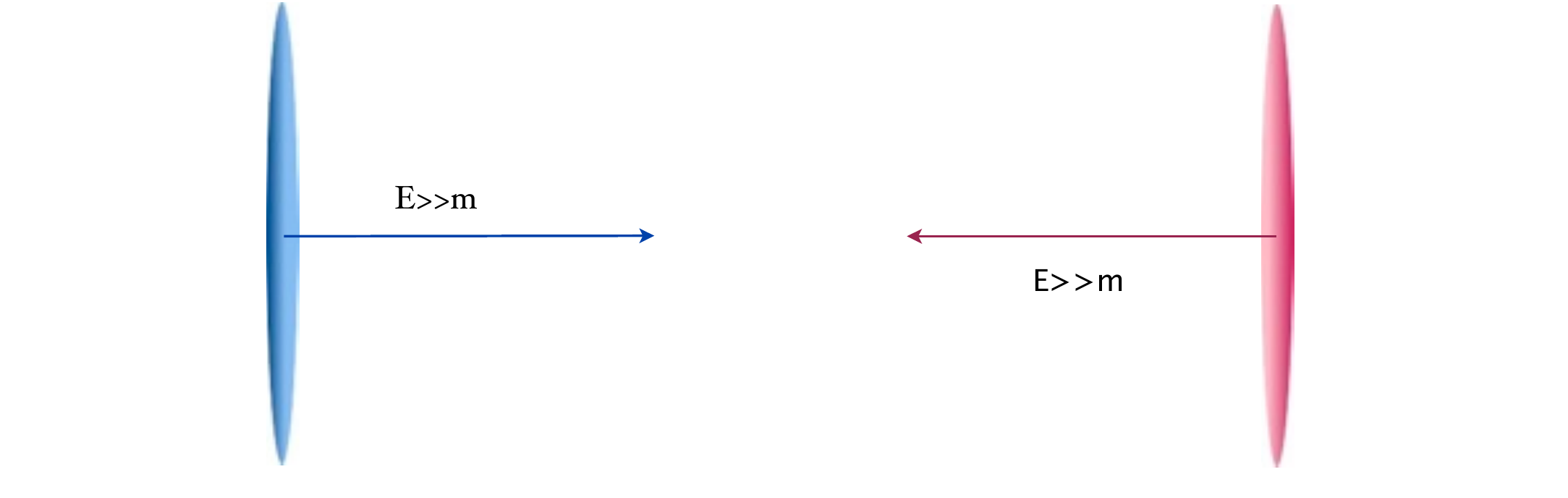}}
\resizebox*{0.70\columnwidth}{!}{\includegraphics{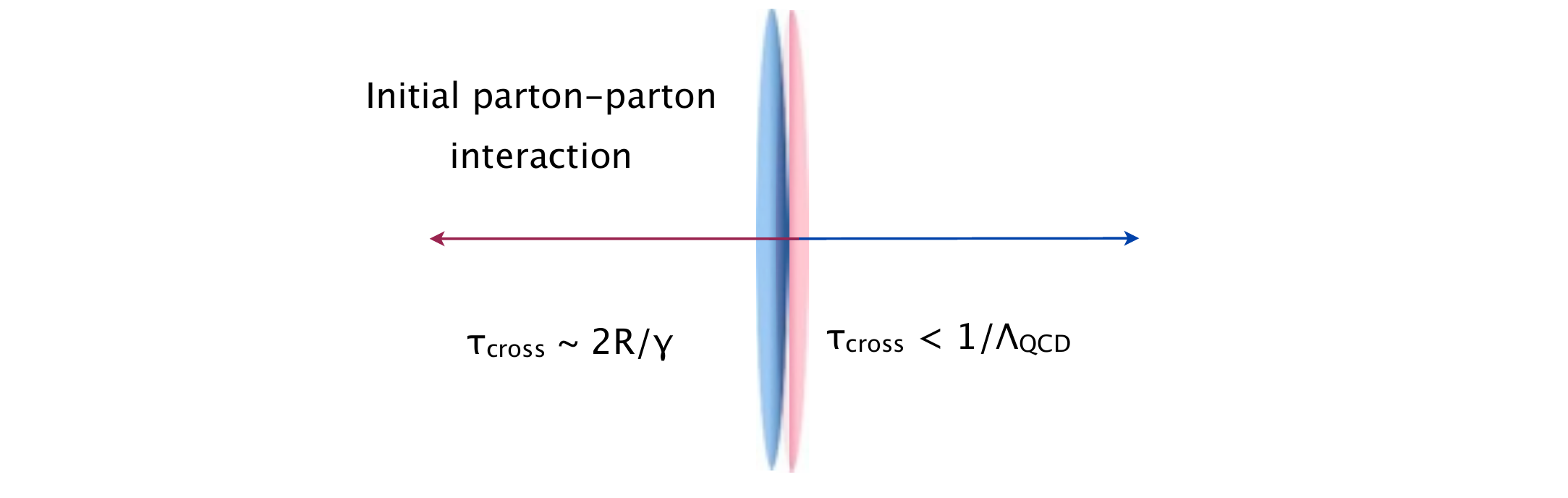}}
\resizebox*{0.70\columnwidth}{!}{\includegraphics{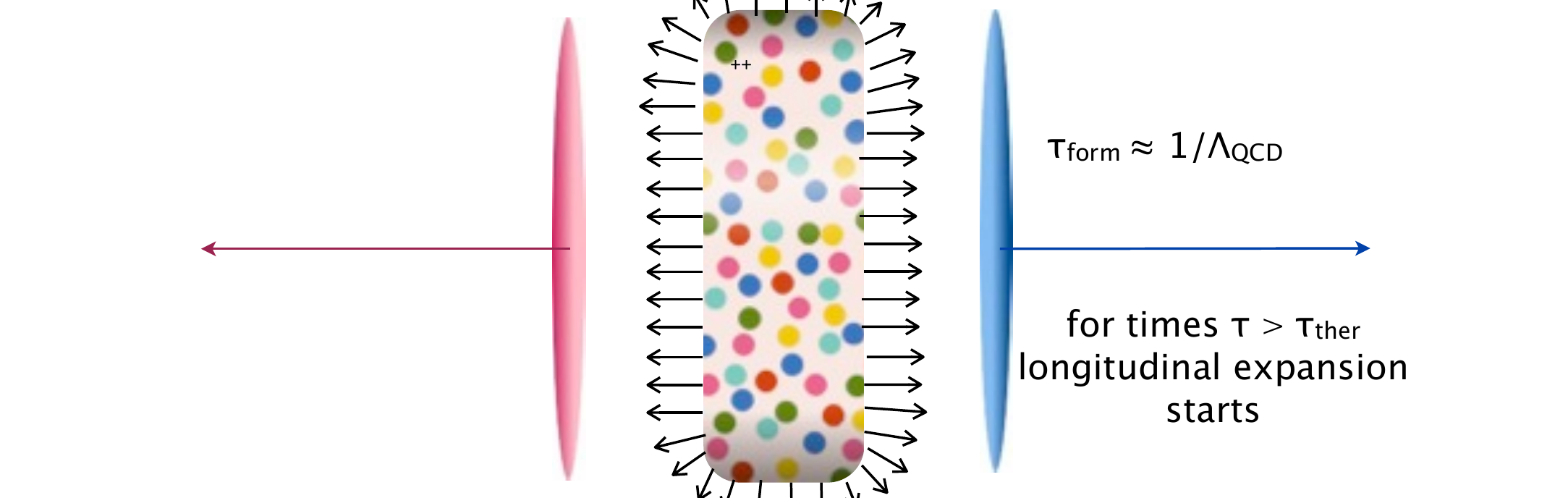}}
\resizebox*{0.70\columnwidth}{!}{\includegraphics{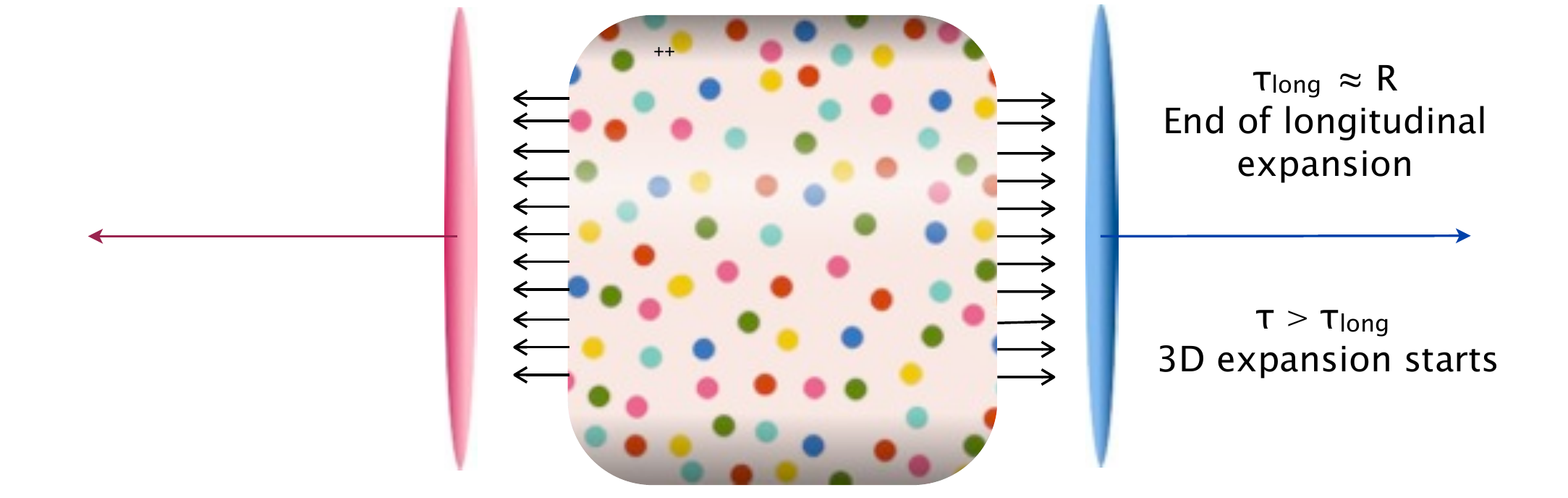}}
\resizebox*{0.70\columnwidth}{!}{\includegraphics{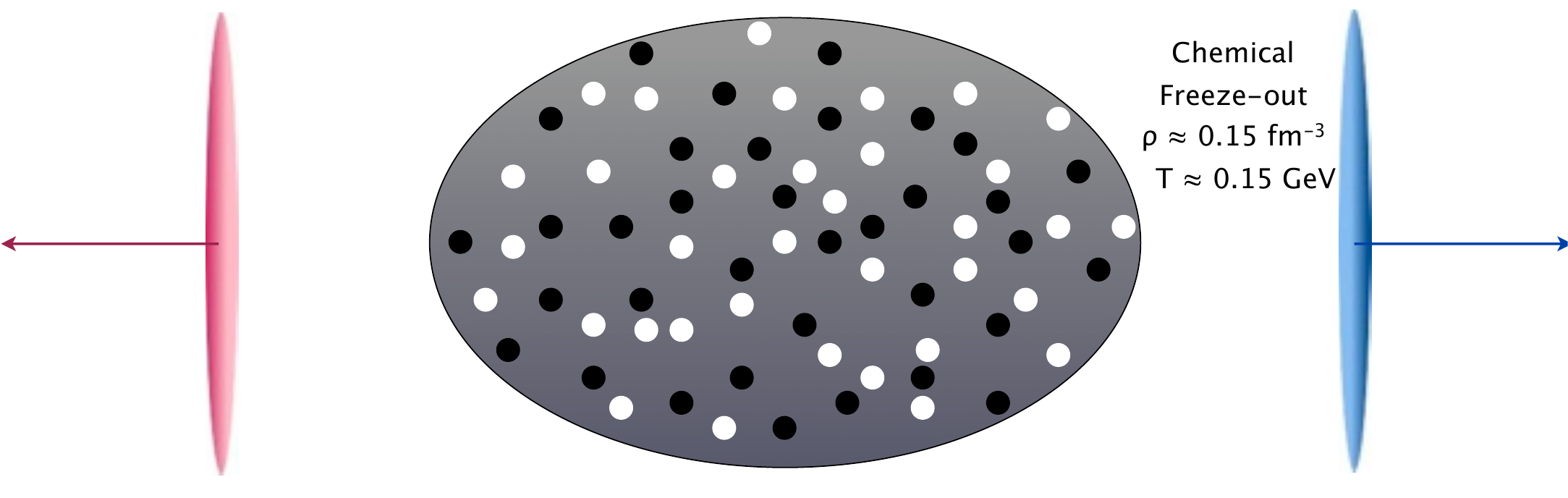}}

\par}
\caption{\label{Coll_BjorkenScenario} 
\emph{Bjorken scenario \protect\cite{Bjor83} for the formation of hot QCD matter. After a formation time $\tau_{\rm form}$ a volume with a high energy density is created. After equilibration at $\tau_{\rm ther}$, the evolution of the hot QCD matter follows the laws of the relativistic hydrodynamics. First, there is a longitudinal expansion until the system reaches a longitudinal size close to its transverse size, then a tridimensional expansion starts until the density is so low that no more inelastic (elastic) collision takes place. The system reaches then the so called chemical (kinetically) freeze-out. Finally all the particles will fly decaying to their daughter particles or reaching the detector. Typically only charged pions, charged kaons, protons, neutrons, photons, electrons and muons will reach the detectors.}
}
\end{figure}

At ultra-relativistic energies, nuclei are seen as pancakes in the center-of-mass system, due to the Lorentz contraction.  The crossing  time of the nuclei can be estimated as 
\begin{equation}
\tau_{cross}=2R/\gamma,
\end{equation} 
where $\gamma$ is the Lorentz factor and $R$ the radius of the nuclei. 

Bjorken assumed the following hypothesis:
\begin{itemize}
\item The crossing time $\tau_{cross}$ is smaller than the time scale of the strong interaction. The latter can be estimated as $\tau_{\rm strong}\sim 1/\Lambda_{QCD}\sim$~1~fm/c. For a nucleus-nucleus collision,  $\tau_{\rm cross} $ is larger than $\tau_{\rm strong}$ only if $\gamma< 12$. That is an energy in the center of mass above $\sqrt{s_{\rm NN}}>$25 GeV (so $E_f>250$ GeV for a fixed-target experiment)\footnote{To be noted that only RHIC and LHC colliders validate this hypothesis with $\gamma$ values of 100 and 1376 (and 2750 after year 2015) respectively.}.
Under this hypothesis, the particles generated by the strong interaction between the nucleon partons, are created once the nuclei have already crossed each other.

\item The distribution of the particle multiplicity as a function of the rapidity is assumed to be uniform. This is partially verified experimentally in  Au-Au collisions at RHIC \cite{Phob05b}, in proton-antiproton collisions at Tevatron  \cite{CDF90}, and at SPS energies \cite{Bjor83}. 
This condition ensures a rapidity symmetry of the system, allowing to create a uniform energy density in different rapidity slices, which simplifies considerably the description of the hydrodynamical evolution of the system.  

\end{itemize}

Let's consider the volume centred in the nucleus crossing plane, at a time $\tau$ after the nucleus crossing. This volume has a cylindrical shape with a thickness $2\Delta d$ along the beam axis direction and a radius $R\sim 1.124A^{1/3}$ in the transverse plane. This volume will contain all the particles produced with a speed along the beam axis (z axis) below $\beta_z \leq \Delta d / \tau$.  Since $\beta_z = \tanh{(y)} \sim y$ for $y \rightarrow 0$, the rapidity range $\Delta y$ around $y=0$ of particles with a $\beta_z \leq  \Delta d / \tau$ will be
\begin{equation}
\Delta y = \frac{2\Delta d}{\tau}.
\end{equation}
and the total energy in the  volume considered will be :
\begin{equation} 
E = \Bigg| \frac{dE}{dy} \Bigg|_{y=0} \times \frac{2\Delta d}{\tau},
\end{equation}
where $dE/dy$ is the total energy created by the strong interaction between the nuclei at $y$=0. For other rapidity domains, the previous expression can be easily generalised replacing the total energy by the transverse energy $E_{\rm T}$. 
Finally, we can calculate the energy density in the volume\footnote{There is a factor 2 difference with respect to equation (3) in the original publication of Bjorken \protect\cite{Bjor83}. It is a known typo error in the original publication.}:
\begin{equation} 
\epsilon(y) = \Bigg| \frac{dE_{\rm T}}{dy} \Bigg| \times \frac{1}{\pi R^2 \tau},
\end{equation}
which links the energy density with the transverse energy produced per unit of rapidity.

\subsubsection{Formation.} 
The initial energy density can then be estimated assuming the time scale needed for the production of particles, as 
$\tau_{\rm form}\sim\tau_{\rm strong}\sim 1~$fm/c \footnote{Other estimates that provide smaller $\tau_{\rm strong}$ in the range 0.2-0.5~fm/$c$ can be foreseen \protect\cite{Phen05}.}.

Bjorken estimated the energy density for heavy ion collisions at beam energies of the Sp$\bar{{\rm p}}$S collider at CERN, that were $\sqrt{s_{\rm NN}} \sim 500$ GeV per nucleon pair\footnote{The energy of the collider Sp$\bar{{\rm p}}$S is close to the available energies at RHIC: 200 GeV per nucleon pair}, and he obtained that the initial energy density were about 2-20 ~GeV/fm$^3$, largely above the critical energy density to form the QGP. One can redo the exercise for heavy ion collisions at Tevatron energies ($\sqrt{s_{\rm NN}} \sim 1.8$ TeV \cite{CDF88,CDF90}), and then the initial energy density would be 4-30 GeV/fm$^3$.

Note that in Fig. \ref{Coll_BjorkenScenario} only hot matter created around mid-rapidity and its evolution is presented. Indeed one should keep in mind that the hot hadronic matter is created in the full rapidity range where the particle density is high enough to reach equilibrium. At RHIC energies, this is about 5 units of rapidity and at LHC energies about 8 units of rapidity. In the laboratory system, the hot matter slices at larger rapidities are indeed narrower due to the Lorentz contraction.

\subsubsection{Thermalisation.} 
The particles produced inside the volume considered will interact.
At these energy densities, and assuming a mean energy $\langle E \rangle$=500 MeV,  $\epsilon/\langle E \rangle \sim$~$8 - 60$ particles per fm$^3$ will be reached.
The  average path length of particles inside the volume can be estimated as $\lambda \sim 0.02 - 0.12$~fm, if one assumes an interaction cross-section of 10 mb. One could hope that the system will thermalise at a time $\tau=\tau_{ther}$.
Note that this is a strong assumption that must be i) validated by the experimental results and ii) supported by theoretical calculations. Experimental results seem to agree with the assumption of a fast thermalization of the system, but the theory has not been able to explain how thermal equilibrium could be reached in such a short time scale. This reminds a fundamental question to be answered and it is still a challenge for QCD theory to describe the first instants of the nucleus-nucleus collision at ultra-relativistic energies. In principle, the initial state of the nucleus-nucleus collision is characterised by the interaction of two high-density gluon clouds. In this respect, classical limits of the QCD theory (like the Colour Glass Condensate \cite{Gelis:2010rs}) seem to be the best theoretical tool to study this problem. The typical Bjorken $x$ of the two gluon clouds is $\langle x \rangle\sim10^{-2}$ at RHIC and $\langle x \rangle\sim10^{-3}$ at the LHC
 
\subsubsection{Longitudinal expansion.}
At stages $\tau \geq \tau_{ther}$ the system should evolve like a fluid, following the laws of the relativistic hydrodynamics. First a longitudinal expansion will take place since the pressure gradient in the beam direction will be larger than that in the transverse plane. It is expected that the energy density will evolve as  $\epsilon\sim1/\tau^n$ with $1\leq n\leq4/3$, which is obtained from the hydrodynamic law  \cite{Bjor83}
\begin{equation}
\frac{d\epsilon}{d\tau} = - \frac{\epsilon+p}{\tau}.
\end{equation}
and for an ideal ultra-relativistic gas, this becomes $\epsilon=3p$ and thus $n$=4/3.
The longitudinal expansion stays as a good approximation for stages  $\tau \leq \tau_{long}\sim R$.

\subsubsection{3D expansion and freeze-out phase.}
For stages $\tau\geq\tau_{long}$ the system will evolve via a 3 dimensional expansion until the freeze-out stage is reached. At freeze-out, particle density is low enough to assume that particles do not interact, travel in the vacuum, can decay and finally reach the detector. Naively, the freeze-out will take place when the average path length  of particles is similar to the size of the system $\lambda \sim R$. For a cross-section of 10 mb, this corresponds to 0.15 particles per fm$^3$ and therefore an energy density of 
\begin{equation}
\epsilon_{gel}~\sim~0.15~{\rm fm}^{-3}~\times~0.5~{\rm GeV}~\sim~0.075~{\rm GeV/fm}^3. 
\end{equation}
It is then expected that the freeze-out takes place as a hadron gas phase.
Note that for a freeze-out temperature of $T_{gel}=150$~MeV,  one gets $\epsilon/T^4 \sim 1.2$, which fits pretty well with the prediction of lattice QCD calculations  of Fig. \ref{Tran_QCDReseau}.
Finally it is worth mentioning that elastic cross-section is larger than inelastic one and one expects to observe two different freeze-out stages: chemical and kinetic freeze-out ones.

\subsection{Heavy ion accelerators and colliders}
Developments in heavy ions beams at ultra-relativistic energies have been performed in parallel as that at intermediate energies, since the main technical limitation was the ion source and the heavy-ion injection at low energies.
The first heavy ion beams at relativistic energies where produced at AGS (BNL, USA) and at SPS (CERN, Switzerland) in the 80's. The energy in the centre of mass was 5 and 18 GeV per nucleon pair, respectively.
The first heavy ion collider was RHIC, built at BNL, which provided the first Au-Au collisions at $\sqrt{s_{\rm NN}}$=130 GeV in June 2000 and reached in 2001 its nominal energy of $\sqrt{s_{\rm NN}}$=200 GeV.
Finally, LHC provided its first heavy ion collisions of Pb beam at $\sqrt{s_{\rm NN}}$=2760 GeV, a 14-fold increasing step with respect to RHIC, in November 2010 and hopefully this will turn into a 28-fold factor (5500 GeV) from 2015 onwards. Today, RHIC and LHC are developing their heavy ion programs which are foreseen until 2025.

\subsubsection{The Alternating Gradient Synchrotron at BNL}
The AGS synchrotron  was built in 1957 and allows the acceleration of high intensity proton beams at 33 GeV. Several Nobel prizes were obtained (1976, 1980 and 1989) linked to discoveries at AGS: J/$\psi$ discovery in 1974, observation of the CP violation of the weak interaction in 1963  and the discovery of the muonic neutrino (1962). Since 1986, the AGS synchrotron has been used to accelerate Si ions at energies of 14 GeV per nucleon, after the construction of the beam line to inject heavy ions in AGS from the Tandem Van de Graaf (built in 1970). The Si beam from the tandem has an energy of 6.6 MeV per nucleon. The construction of the AGS booster in 1991 allowed to increase the AGS beam intensity and to accelerate heavier ions like Au up to 11 GeV per nucleon. Negative Au$^-$ ions are extracted from the source and accelerated by the tandem to 1.17 MeV per nucleon and stripped to a beam of Au$^{+32}$. This beam is then injected in the AGS booster where the Au ions are accelerated to 90 MeV/nucleon. Finally the Au beam is stripped and injected into AGS where it is accelerated to the nominal energy of 11 GeV per nucleon. For 14 years, several fixed target heavy-ion experiments took place, like  E866, E877, E891, E895, E896, E910, E917 to study the hadronic matter at high temperature\footnote{Note that it is not clear the AGS could form deconfined matter since the initial energy density could be below 1 GeV/fm$^3$.}.

\subsubsection{The Super Proton Synchrotron at CERN}
The SPS  was built in 1976, allowing for proton acceleration until 500 GeV.
First, protons are accelerated by a linear accelerator called LINAC2, and then injected into the booster of the PS (Proton Synchrotron) and finally they are injected into the  SPS to reach their nominal energy of 500 GeV\footnote{Initially the SPS was a proton accelerator. But SPS became a proton-antiproton collider with to the additional injection of antiproton beam. The latter was attainable thanks to the stochastic-cooling technique in the SPS ring. The first collisions  $p\bar{p}$ in SPS took place in 1981 at a center of mass energy of 520 GeV. Two years later, the electroweak bosons were discovered by the UA1 and UA2 experiments. The stochastic-cooling and the discovery of the $W$, and $Z$ bosons was awarded with the Nobel prize of physics in 1984.}. From 1986, the new electron-cyclotron resonance (ECR) ion source allowed the injection of multi charged heavy ions in the CERN accelerator system (LINAC3, PS booster, PS and SPS).  The beam leaving from an ECR ion source containing a Pb plasma, has an energy of 2.5 KeV per nucleon with an ion charge $Q=+27$, and they are injected in the LINACS3 linear accelerator  reaching a beam energy of 4.2 MeV per nucleon. Then the beam is stripped via a thin C layer 1 $\mu$m  thick, and becomes a $^{+53}$Pb beam, which is injected in the PS booster and PS accelerator, reaching an energy of 4.25 GeV per nucleon. The Pb ions are then fully stripped in an aluminium layer of 1 mm thick, and they are injected in SPS to reach an energy of 158 GeV per nucleon. This beam is finally directed to the experimental fixed-target halls in SPS north area (NA) in France or SPS west area (WA) in Switzerland, where heavy ion collisions at $\sqrt{s_{\rm NN}}$ take place. In addition to Pb, other ions have been also accelerated at SPS. At the beginning of the SPS heavy ion program, beams of O and S were accelerated at energies about 60 and 200 GeV per nucleon, and in the last days In ions were used for the NA60 experiment. During 20 years, many heavy ion experiments were built, installed and contributed to the SPS heavy-ion physics programme: WA80, WA93, WA98, WA85, WA94, WA97, NA57, Helios-2, NA44, CERES, Helios-3, NA35, NA49, NA36, NA52, NA38, NA50 et NA60. In 2000, the analysis and interpretation of the obtained experimental results was almost finished and a CERN press released was organised\footnote{http://press.web.cern.ch/press/PressReleases/Releases2000/PR01.00EQuarkGluonMatter.html.}.  They announced that the physical results of the heavy ion fixed-target SPS experiment  NA44, NA45, NA49, NA50, NA52, WA97 / NA57 and WA98 hinted at the existence of a new state of matter in which quarks, instead of being bound up into more complex particles such as protons and neutrons, were liberated to roam freely.
 
\subsubsection{The Relativistic Heavy Ion Collider at BNL}

The first  Au-Au collisions at 130 GeV per nucleon pair took place in June 2000 in RHIC at BNL (USA). It was the first collider ever built for heavy ions. AGS is the injector of RHIC, via a two Au beams at 9 GeV per nucleon which circulate in two different rings  in opposite directions. In RHIC collider, 60 beam bunches in each ring are accelerated to the nominal energy of 100 GeV per nucleon and stored in two rings of 3.85 km perimeter length. The bunches of the two beams can collide in 4 interaction points along the RHIC ring, reaching nominal luminosities  about $2 \cdot 10^{26}~$cm$^{-2}~$s$^{-1}$, that is a Au-Au collisions rate of 800 Hz. Recently, RHIC has been upgraded and is able to provide 5-10 times more instantaneous luminosity. In addition, RHIC collider allows to study collisions of polarised protons at 500 GeV, and collisions of d-Au, Cu-Cu, Au-Au and U-U in the energy range 20-200 GeV per nucleon pair. Since 2000, the four experiments at RHIC: STAR, PHENIX, PHOBOS and BRAHMS have developed a high quality physics program, producing a huge amount of experimental results. Today only the two major experiments: PHENIX and STAR are still active and taking data.

\subsubsection{The Large Hadron Collider at CERN}
The LHC at CERN uses SPS as injector. SPS was upgraded to generate a Pb ions beam at 177 GeV per nucleon, that are accelerated to a beam energy of 1.38 TeV.  LHC provided the first Pb-Pb collisions at 2.76 TeV in November 2010, increasing by a factor 14 the centre-of-mass energy at RHIC. In November 2011 a new heavy-ion run took place at the same energy and the nominal instantaneous luminosity was reached, $\sim 5 \cdot 10^{26}~$cm$^{-2}~$s$^{-1}$. It is expected that the nominal energy, 5.5 TeV per nucleon pair,  will be reached after the long shutdown during 2013-2014. In principle the instantaneous luminosity at LHC and beam lifetime is limited by the huge cross-section of i) electromagnetic production of electron-positron pairs where the electron is captured by the Pb ions and ii) electromagnetic excitation of the Pb nucleus giant resonance, leading to neutron emission. Both processes are responsible for the Pb beam loss at LHC energy.  LHC will be upgraded in 2018 to increase by a factor 10 the instantaneous luminosity of the Pb-Pb collisions. At LHC, three of the four LHC experiments participate in the heavy ion program:  ALICE, ATLAS and CMS. ALICE is the only LHC experiment devoted to the study of QGP. LHC will provide the first proton-Pb collisions at the beginning of 2013.

\section{Some bases about collision centrality and the nuclear modification factor}

In Fig. \ref{Coll_BjorkenScenario}, the Bjorken scenario is presented for a central (zero impact parameter, $b$) collision is presented. Actually, collisions at any impact parameter between $b=0$ and $b=R_1+R_2$ (the sum of the nuclear radius) could occur in the laboratory. It turns out that most of the collisions are indeed peripheral collisions, since the probability density is proportional to $b$. In experiments, the centrality of the collision can be estimated on an event-by-event basis via any observable $\cal{C}$ that monotonically varies with the impact parameter of the collision.  The observable $\cal{C}$ can be the charged particle multiplicity or transverse energy in a given pseudo-rapidity interval, or energy at zero degree (at rapidities close to the beam rapidity), etc... Let us assume that i) $f(\cal{C})$ represents the distribution of the observable $\cal{C}$ for a sample of non biased nucleus-nucleus collisions, that  ii) $\cal{C}$(b)$\geq$0 and that iii) $\cal{C}$(b=0)~=~0. 
The centrality class $n$\% of the most central collisions consists of nucleus-nucleus collisions where the observable 
${\cal{C}} \in (0,{\cal{C}}_n)$ and 
\begin{equation}
n = 100 \times  \frac
{\int_0^{{\cal{C}} _n} f({\cal{C}}) d{\cal{C}} }
{\int_0^\infty f({\cal{C}}) d{\cal{C}} }
\end{equation}
The $n$\% most central collisions are usually referred to as the centrality class 0-$n$\%. Therefore the reaction class $m$\%-$n$\% ($m<n$) is defined by the collisions where the observable  ${\cal{C}} \in ({\cal{C}}_m,{\cal{C}}_n)$.

One of the experimental methods to quantify the nuclear medium effects in the production of a given observable (Ob) is the measurement of the nuclear modification factor ($R^{\rm Ob}_{\rm AA}$) in nucleus-nucleus (A-A) collisions, defined as:
\begin{equation}
R^{\rm Ob}_{\rm AA}  = \frac{ Y^{\rm Ob}_{\rm AA} }{\langle N_{coll} \rangle \, \, Y^{\rm Ob}_{\rm pp}}
\end{equation}
where $\langle N_{coll} \rangle$ is the average number of binary nucleon-nucleon collisions\footnote{ The average number of binary nucleon-nucleon collisions can be estimated by the product of the average nuclear overlap function (of the nucleus-nucleus collision) and the inelastic proton-proton cross section~\cite{Mill07, Ente03}.} and  $Y^{\rm Ob}_{\rm AA}$ ($Y^{\rm Ob}_{\rm pp}$) is the invariant yield of the observable Ob in A-A (pp) collisions at a given (same) center-of-mass energy.  In the absence of nuclear matter effects, the nuclear modification factor should be equal to unity for experimental observables commonly called \emph{hard probes} (large p$_{\rm T}$ particles, jets, heavy-flavour, etc). 
A similar factor $R^{\rm Ob}_{\rm pA}$, measured in p-A collisions, is crucial in order to disentangle hot and cold nuclear matter effects in A-A collisions.

\section{Brief summary of the experimental results at RHIC and at the LHC}

Due to a lack of time, I have not been able to complete satisfactorily these proceedings. For this reason, I am giving here a brief summary of the main results from RHIC (12 years of heavy ion programme) and from LHC (after the two first years of heavy ion programme). 

\subsection{Initial energy density}

\begin{figure}
{\centering 
\resizebox*{0.80\columnwidth}{!}{\includegraphics{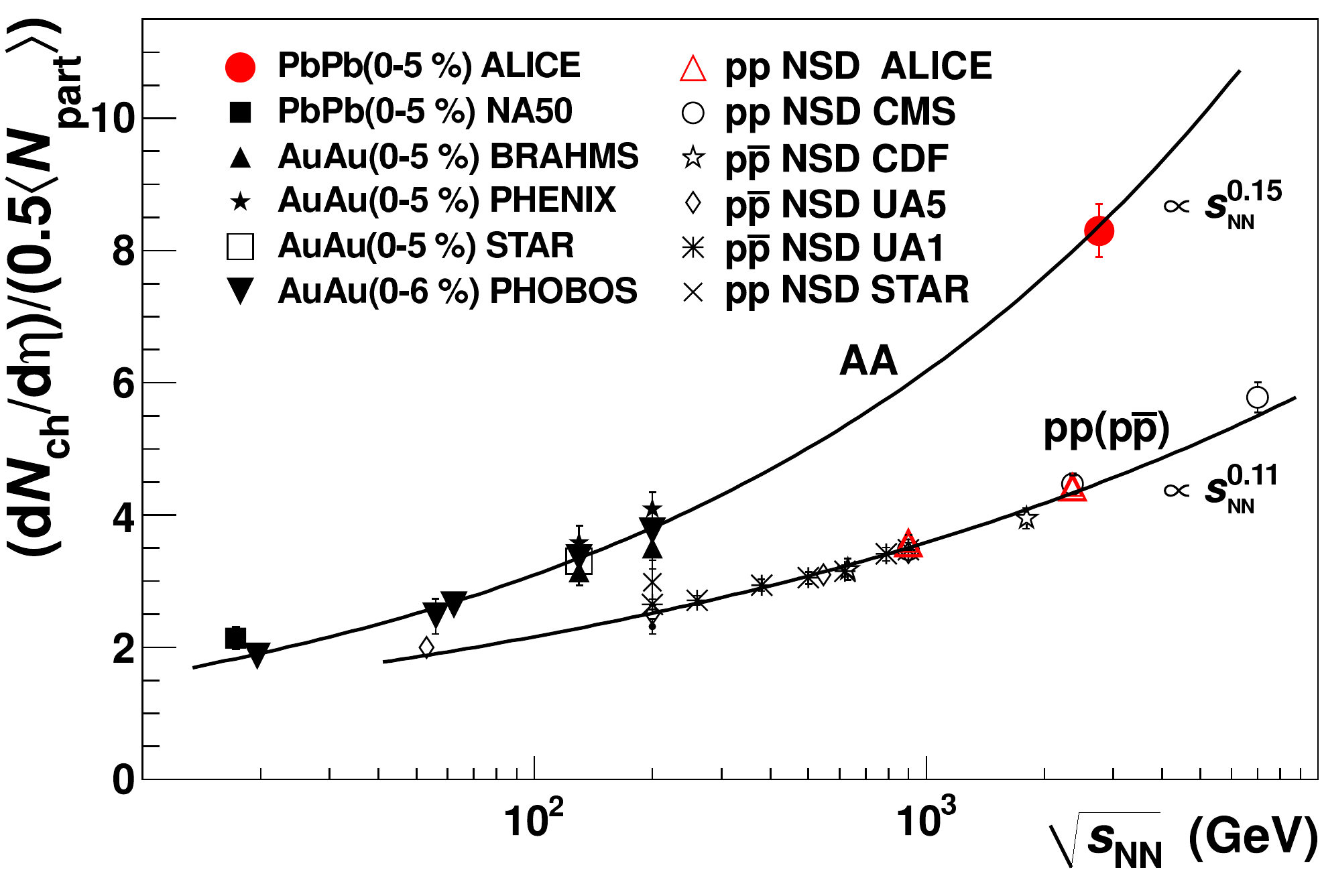}}
\par}
\caption{\label{1011-3916v3_Fig3}
\emph{Charged particle pseudo-rapidity density per participant pair for central nucleus-nucleus collisions. The solid lines $\propto s^{0.15}_{\rm NN}$ and $\propto s^{0.11}_{\rm NN}$ are superimposed on the diffractive pp collisions as a function of $\sqrt{s_{\rm NN}}$ heavy-ion and pp data, respectively. Figure 3 in reference \cite{Alic10a}.}}
\end{figure}

The multiplicity of charged particles $dN_{ch}/d\eta$ was measured at RHIC \cite{Phob00, Phob02,Phen05, Phen05d}  (as well as the transverse energy \cite{Phen01b, Star04}). The most central Au-Au collisions at 200 GeV generate more than 600 charged particles per unit of pseudo-rapidity at mid-rapidity, which should correspond to about 900 (charged and neutral) particles per unit of pseudo-rapidity\footnote{Grosso-modo one assumes that the pions are the most abundant produced particles. Only the $\pi^\circ$ is neutral, so the total number of particles per unit of rapidity can be estimated as $600\times3/2$.}. Using the Bjorken model one can estimate that the initial energy density at mid-rapidity amounts to about 5-15 GeV/fm$^3$. In addition the charged particle multiplicity remains constant within $\sim$10\% for 5 units of pseudo-rapidity ($|\eta| \lesssim$ 2.5) \cite{Phob11}. In the most central Pb-Pb collisions at 2.76 TeV, CMS has measured that transverse energy at mid-rapidity is about $\sim$2 TeV per unit of pseudo-rapidity \cite{CMS12}. ALICE and ATLAS measured about $\sim$1600 charged particles per unit of pseudo-rapidity \cite{Alic10a, Alic10b,Atla11} (see Fig. \ref{1011-3916v3_Fig3}). Considering the increase of the mean hadron p$_{\rm T}$ at the LHC, the initial energy density at LHC is about three times larger than in Au-Au at RHIC top energy: 15-30 GeV/$\rm fm^3$. In both cases, the energy densities are several times larger than the critical energy density to form deconfined matter. Assuming that the system quickly equilibrates, the initial temperature could be estimated from the lattice QCD, assuming $\mu_B\sim 0$, (see Fig. \ref{Tran_QCDReseau}) as 
\begin{equation}
{\rm T}^4 [{\rm MeV}^4] \approx \frac{ 200^3 \times 10^3}{12.5} \times \epsilon ~~[{\rm GeV/fm}^3] 
\end{equation}
which gives an estimate of the initial temperature of 240-320 MeV at RHIC top energy and 310-370 MeV at LHC 2.76 TeV energy.

\subsection{Equilibration}

\begin{figure}
{\centering 
\resizebox*{0.45\columnwidth}{!}{\includegraphics{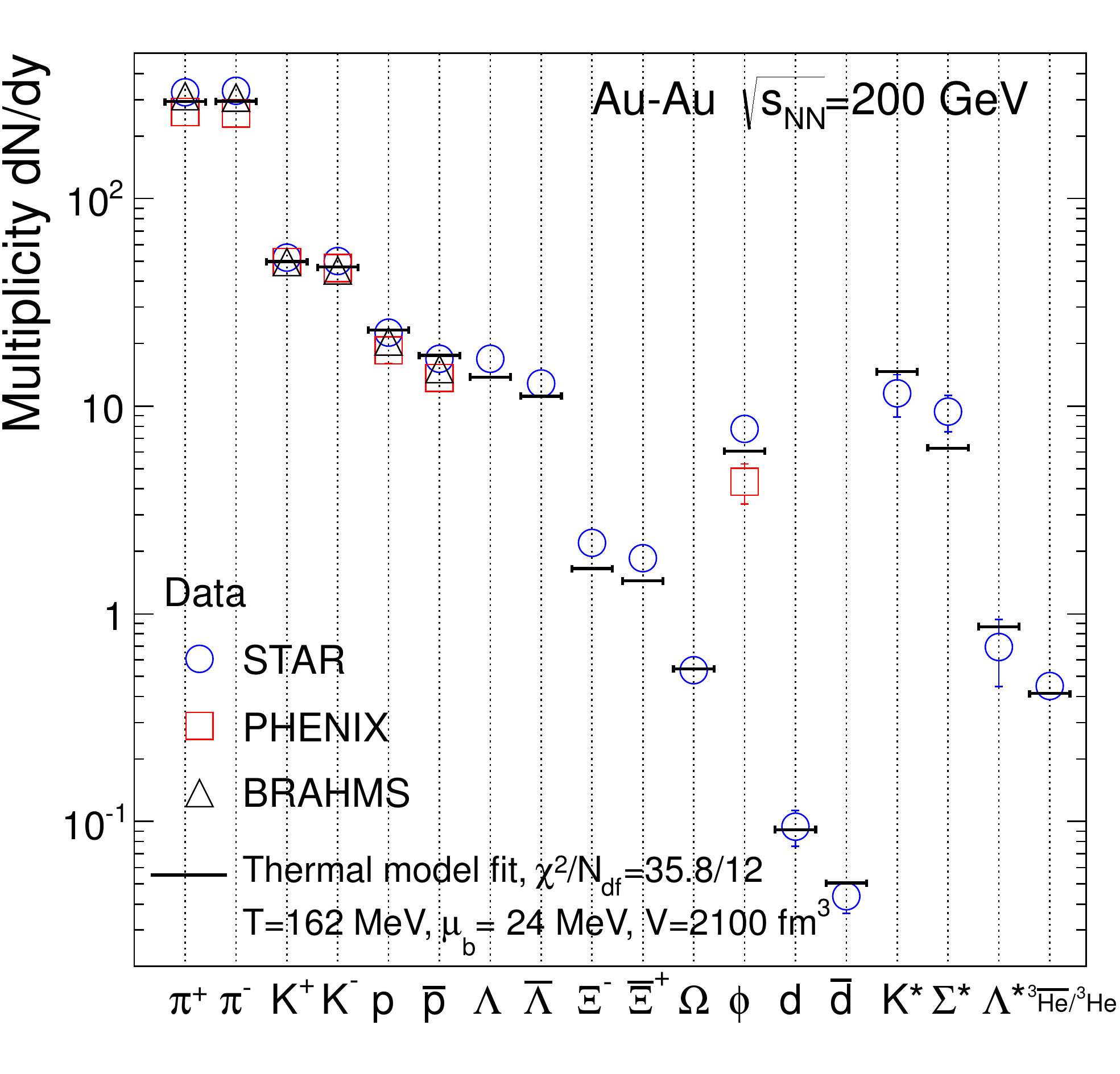}}
\resizebox*{0.45\columnwidth}{!}{\includegraphics{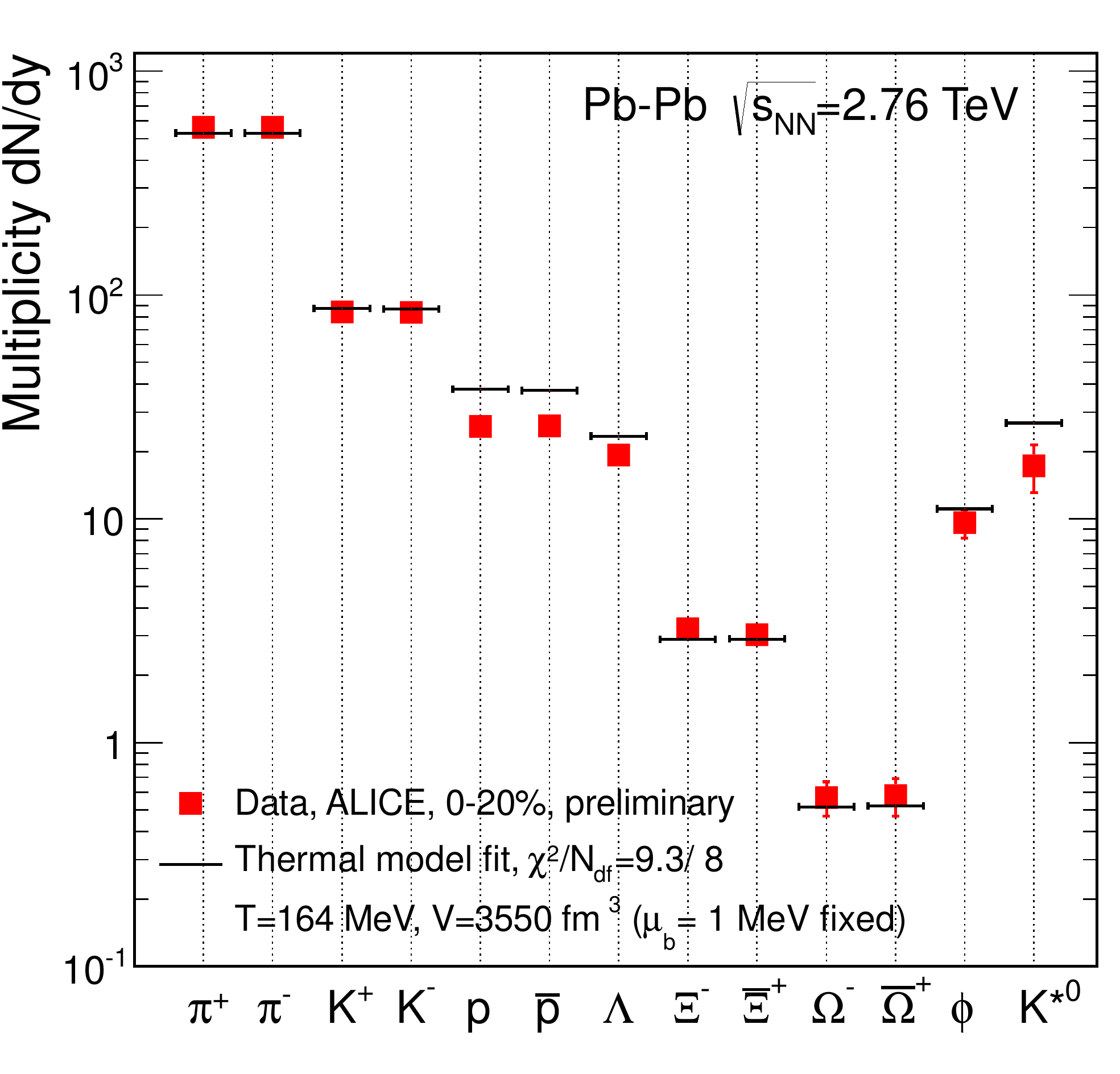}}

\par}
\caption{\label{1210-7724v1}
\emph{ Left: Comparison of thermal model predictions with RHIC data. Right: Thermal model fits to ALICE data on hadron production in central Pb--Pb collisions. From reference \cite{Andr12}.}}
\end{figure}

Integrated hadron yields at mid-rapidities were studied in both energy domains. It is observed that the hadron yield ratios can be successfully described by a statistical model \cite{Andr04}. In this model, the expanding hot system hadronizes statistically at the freeze-out, and therefore the hadron yields are given by the following expression:
\begin{equation}
n_i = \frac{N_i}{V} = \frac{g_i}{2\pi^2}\int_0^\infty \frac{p^2dp}{\exp[(E_i-\mu_i)/T] \pm 1}
\end{equation}
with (+) for  fermions and (-) for bosons, $T$ is the temperature, $N_i$ is the total number of hadrons of the species $i$, $V$ the total volume of the system,  $g_i$ is the isospin and spin degeneration  factor, 
$E_i$ the total hadron energy and $\mu_i$ the chemical potential. Considering zero total strangeness and isospin of the system, one can consider $\mu_i=\mu_b$ where $\mu_b$ is the baryonic chemical potential.
Therefore only two parameters are needed to predict the hadron yield ratios: the freeze-out temperature  and the baryonic potential. The analysis of hadron yield ratios allows to extract a similar freeze-out temperature of $\sim$ 160 MeV at RHIC and at the LHC (see Fig. \ref{1210-7724v1}). The baryonic potential is $\mu_b\sim$20 MeV at RHIC and, as expected, a lower $\mu_b$ at LHC, indeed close to zero \cite{Andr09, Andr12}. The value of the temperature at chemical freeze-out is indeed very close to the phase transition temperature as predicted by lattice calculations presented in section \ref{lattice}.  One should notice that, at LHC energies, proton and antiproton yields normalised to the pion yields exhibit an anomalous behaviour that has to be further investigated  \cite{Andr12}.

\begin{figure}
{\centering 
\resizebox*{0.80\columnwidth}{!}{\includegraphics{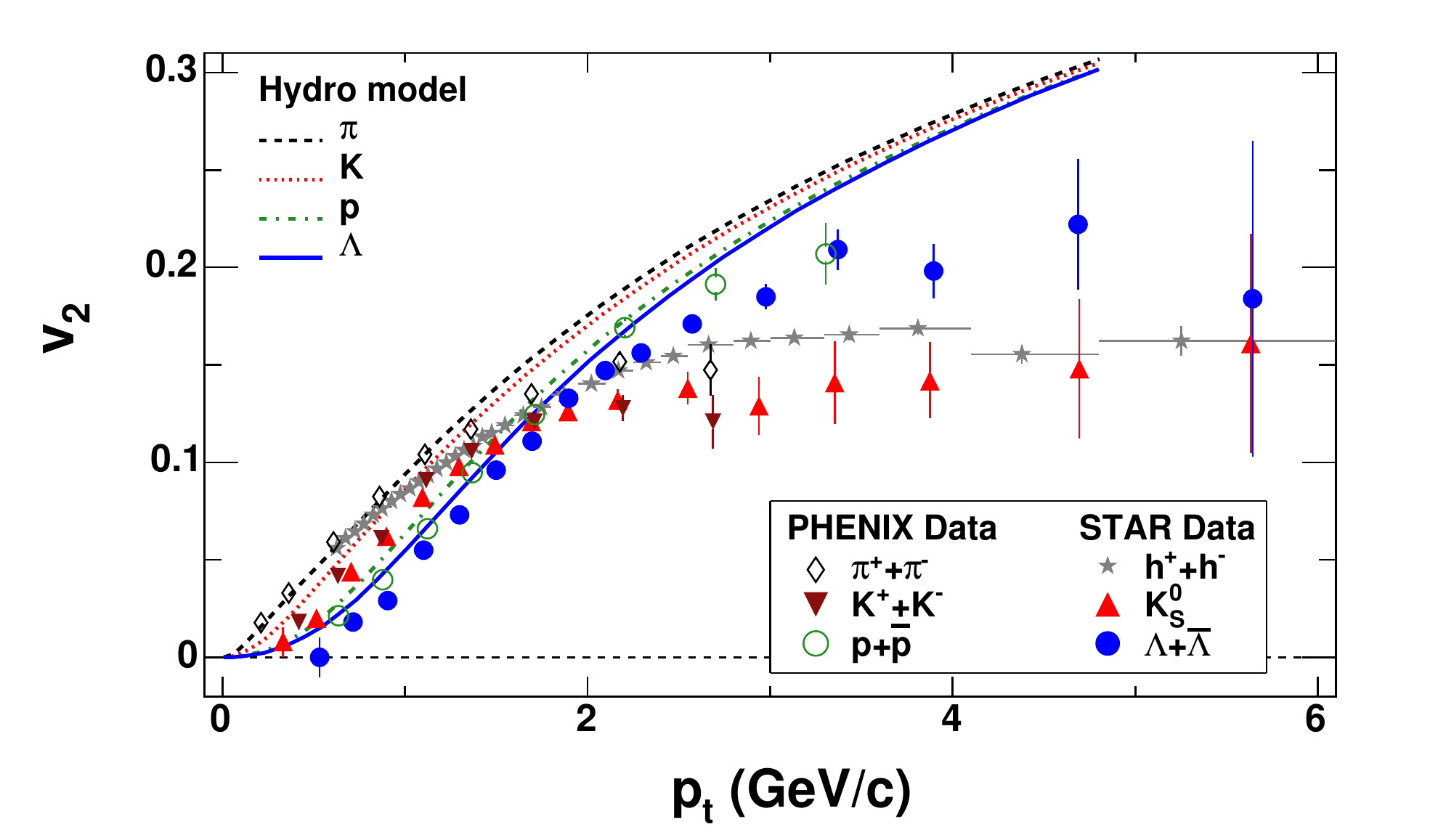}}
\par}
\caption{\label{0409033v3}
\emph{The elliptic flow as a function of the transverse momentum measured by PHENIX and STAR collaborations for hadrons, pions, kaons, protons and Lambda baryons. Figure 10 in reference \cite{Star05b}. }}
\end{figure}

The azimuthal distribution of particles in the plane perpendicular to the beam direction is an experimental observable which is also sensitive to the dynamics of the early stages of heavy-ion collisions. When nuclei collide at finite impact parameter (non-central collisions), the geometrical overlap region and therefore the initial  matter distribution is anisotropic (almond shaped). If the matter is strongly interacting, this spatial asymmetry is converted via multiple collisions into an anisotropic momentum distribution \cite{Ollitrault:1993ba}. The second moment of the final state hadron azimuthal distribution with respect to the reaction plane is called the elliptic flow ($v_2$):
\begin{equation}
E \frac{d^3N}{d^3 \vec{p}} =  \frac{1}{2\pi}\frac{d^2N}{p_T dp_T dy} \Biggl[ 1 + \sum_{n=1}^{\infty} \Bigl\{ 2 v_n \cos{[n(\phi-\Psi_R)]} \Bigr\}  \Biggr]
\end{equation}
where $\Psi_R$ is the reaction plane, defined by the beam axis and the impact parameter.

The elliptic flow has extensively been studied at RHIC \cite{Star01, Star05b, Phob11b, Phen09, Star09, Star11} (see Fig. \ref{0409033v3}), and recently at LHC energies \cite{Alic10c, Atla12, CMS12b, Alic11, Atla12b}.  Indeed the predictions from hydrodynamical models explain quite well most of the measurements of the elliptic flow of light hadrons at low p$_{\rm T}$ (p$_{\rm T}< 2-3$ GeV). The elliptic flow measurements have been one of the major observations at RHIC, evidencing that : i) the created matter equilibrates in an early stage of the collision,  and then it evolves following the laws of the hydrodynamics; and ii) the formed matter behaves like a perfect fluid \cite{Phen05,Star05,Phob05,Brah05}.  Furthermore, several works (see for instance reference \cite{Nagle:2009ip}) managed to extract values of transport properties, like the ratio of the shear viscosity over entropy from the experimental results. The conclusion was that the hot matter behaves as a perfect fluid  and the mean free path of the constituents is close to the quantum limit. ALICE  presented the first elliptic flow measurement at the LHC \cite{Alic10c} in agreement with other LHC results \cite{Atla12}. It was observed a similarity between RHIC and the LHC of p$_{\rm T}$-differential elliptic flow at low p$_{\rm T}$, which is consistent with predictions of hydrodynamic models (p$_{\rm T} \lesssim 2-3$ GeV/c). The elliptic flow is now being studied in much more details at LHC, for identified hadrons and as a function of the pseudo-rapidity (see for instance \cite{Atla12b}). Preliminary results on these topics are intriguing and it is certainly too early to conclude about their interpretation. In addition, the elliptic flow is being studied for the first time at very high p$_{\rm T}$, (up to p$_{\rm T} \sim 40-60$ GeV/$c$) \cite{CMS12b}. Other moments of the hadron azimuthal distribution have been studied at RHIC and at the LHC \cite{Alic11, Atla12}. In particular non-zero $v_3$, was observed and it results from the fluctuations of the initial spatial distribution of the energy density. Complementarily to the $v_2$, the measurement of higher harmonics of the azimuthal distribution is also crucial to constrain the shear viscosity over entropy ratio in models.

\subsection{Initial temperature}

\begin{figure}
{\centering 
\resizebox*{0.80\columnwidth}{!}{\includegraphics{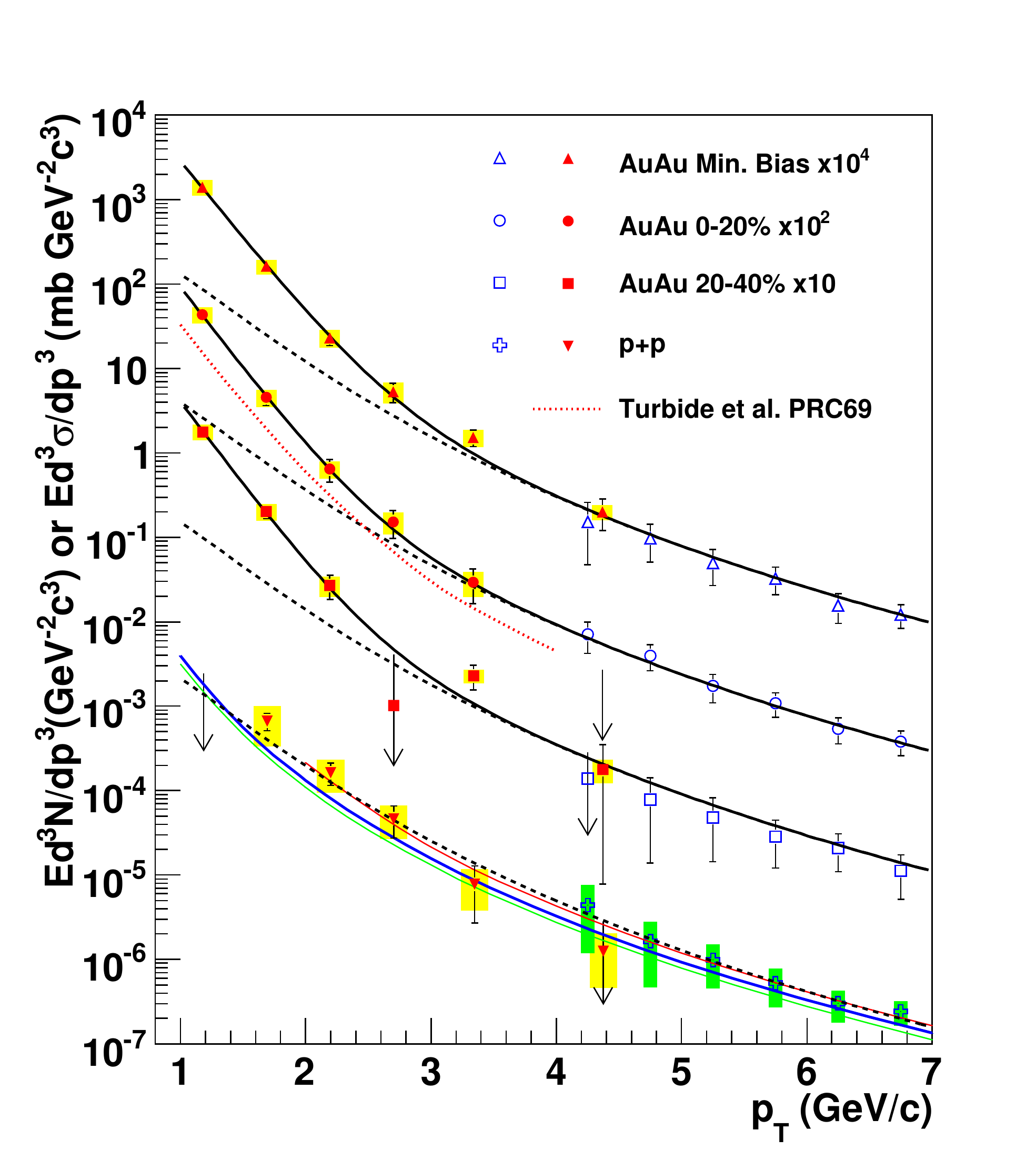}}
\par}
\caption{\label{0804-4168v2}
\emph{Invariant cross section ($pp$) and invariant yield (Au-Au) of direct photons as a function of p$_{\rm T}$.  The three curves on the $pp$ data represent NLO pQCD  calculations, and the dashed curves show a modified power-law fit to the $pp$ data, scaled by $T_{\rm AA}$.  The dashed (black) curves are exponential plus the $T_{\rm AA}$ scaled $pp$ fit.   The dotted (red) curve near the 0-20\% centrality data is a theory calculation. Figure 3 in reference \cite{Phen10}.
}}
\end{figure}

As we have seen in section \ref{thermalradiation}, if QGP drop is formed, it should emit thermal radiation in the high energy $\gamma$ domain. PHENIX collaboration have measured $e^{+}e^{-}$ pairs with invariant masses below 300 MeV/$c^2$ and 1$\leq$p$_{\rm T}\leq$ 5 GeV/$c$ in Au-Au collisions at 200 GeV \cite{Phen10}. The most central Au-Au collisions show a large excess of the dielectron yield (see Fig. \ref{0804-4168v2}). By treating the excess as internal conversion of direct photons, the direct photon yield is deduced. The yield cannot be explained by Glauber scaled NLO pQCD calculations.
However, hydrodynamical models with an initial temperature of 300-600 MeV are in qualitative agreement with the data.The evidence for the production of \emph{thermal} direct photons, with an initial temperature source above the QGP transition temperature represent an important experimental observation. Preliminary results from ALICE about thermal photon production in central Pb-Pb at 2.76 TeV are already available \cite{Alic12}. The supposed thermal photon yield exhibits a 40\% larger inverse slope at LHC than that at RHIC. The latter is in qualitatively good agreement with the expected relative increase of the initial temperature from RHIC to LHC energies.

Quarkonium was proposed as a probe of the QCD matter formed in relativistic heavy-ion collisions more than two decades ago. A familiar prediction, quarkonium suppression due to colour-screening of the heavy-quark potential in deconfined QCD matter \cite{Satz86}, has been experimentally searched for at the SPS and RHIC heavy-ion facilities. 

\begin{figure}
{\centering 
\resizebox*{0.48\columnwidth}{!}{\includegraphics{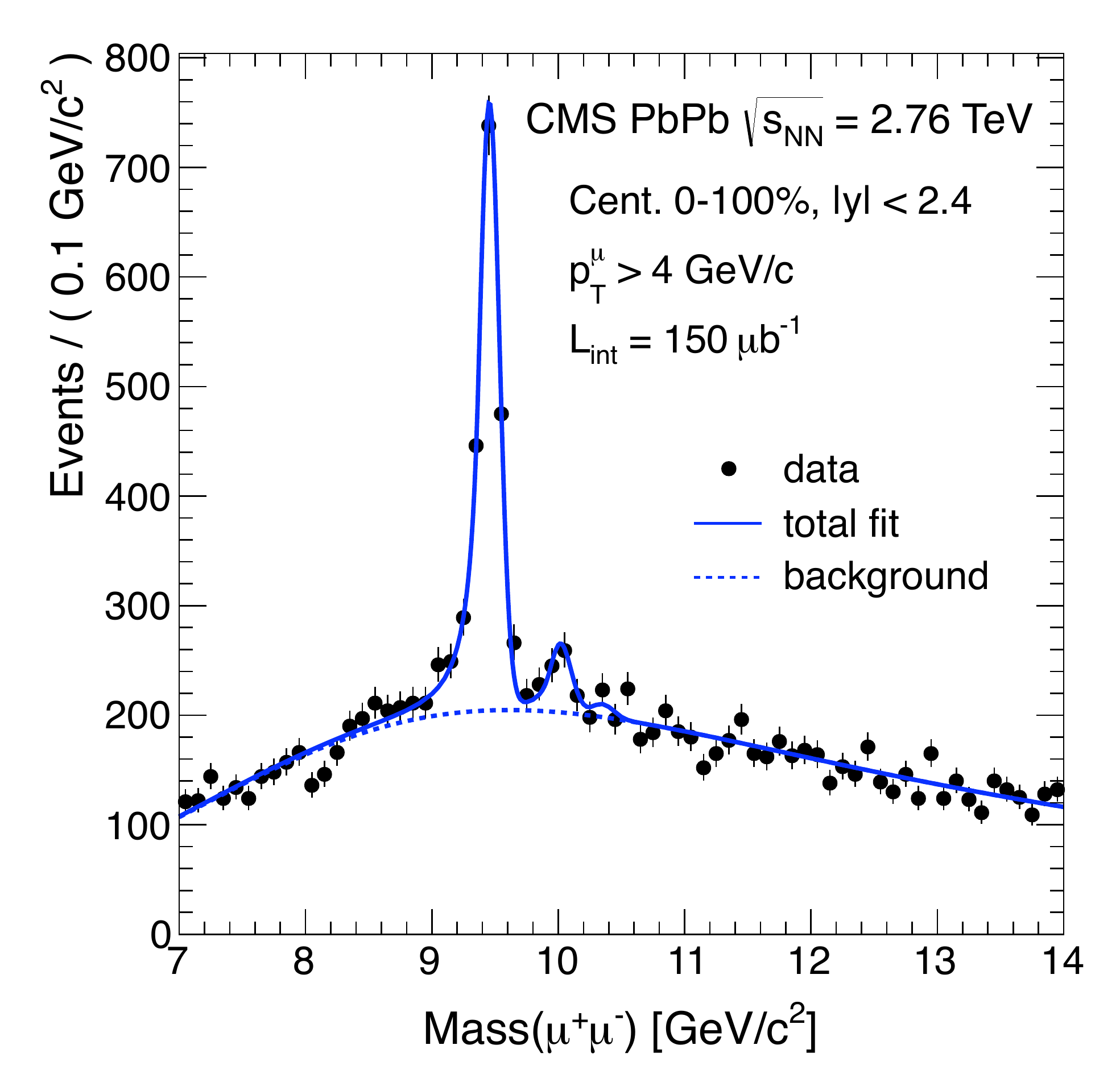}}
\resizebox*{0.48\columnwidth}{!}{\includegraphics{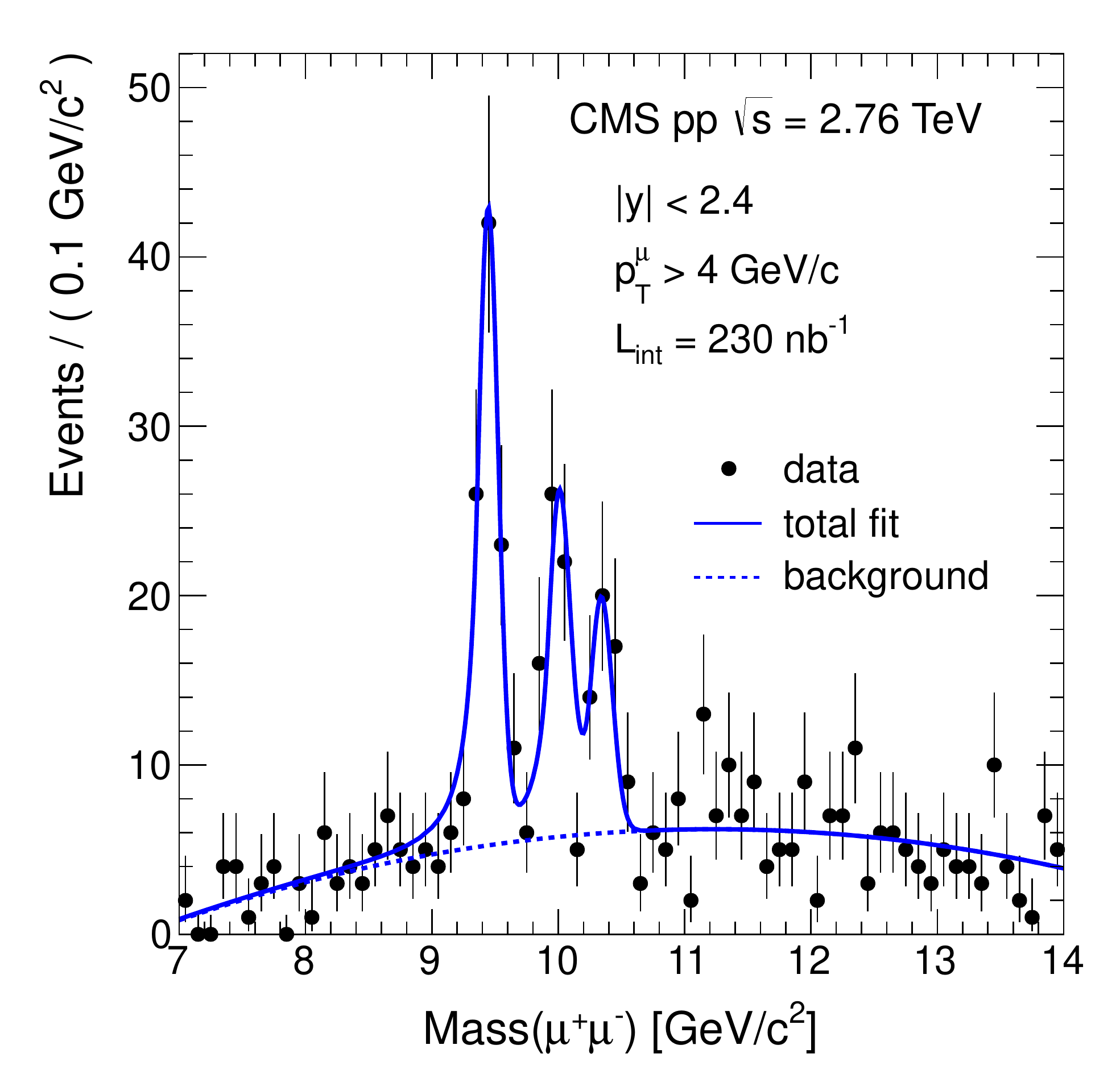}}
\par}
\caption{\label{1208-2826v1}
\emph{Dimuon invariant-mass distributions in Pb-Pb (left) and pp (right) data at $\sqrt{s_{\rm NN}}$= 2.76 TeV. The solid (signal + background) and dashed (background-only) curves show the results of the simultaneous fit to the two datasets. Figure 1 in reference \cite{CMS12d}.  }}
\end{figure}

CMS collaboration has performed the first measurement of the upsilon resonances ($\Upsilon$(1S), $\Upsilon$(2S) and $\Upsilon$(3S)) at the LHC \cite{CMS11, CMS12c, CMS12d}.
The results indicate a significant decrease of the  $\Upsilon$(2S) and $\Upsilon$(3S) R$_{\rm AA}$ (see Fig. \ref{1208-2826v1}). 
The  $\Upsilon$(1S) R$_{\rm AA}$ is about 0.41 for the most central collisions.
One should note that about 50\% of the upsilon production in hadronic collisions is expected to result from the radiative decays of higher bottomonium resonances \cite{Bedj04}. If one assumes that high resonances are dissolved, one would expect to measure a nuclear modification factor for the $\Upsilon$(1S) about 0.5. The present measurement would be compatible with a formation of a QGP at the LHC at an initial temperature between 1.2-2.0 times the critical temperatures (see Tab. \ref{Tab:DissoTemp}), so absolute temperatures between 200-400 MeV. Since the melting temperature of $\Upsilon(2S)$ and J/$\psi$ are expected to be similar (see Tab. \ref{Tab:DissoTemp}) one should expect a similar decrease of the J/$\psi$ R$_{\rm AA}$ at LHC energies.

The PHENIX experiment at RHIC reported the observation of J/$\psi$ suppression in central Au-Au collisions at $\sqrt{s_{\rm NN}}$=200 GeV (10 times higher than the maximum energy in the CM at SPS) \cite{Phen07, Phen11, Phen12}.  Deuteron-gold collisions have been used to constrain cold nuclear matter (CNM) effects at RHIC energies \cite{Phen11b}.   As a consequence, J/$\psi$ suppression due to dissociation in QGP matter is roughly estimated to be 40-80\% in central Au-Au collisions at RHIC energies. Since about 40\% of the J/$\psi$ yield results from the decays of higher resonances, it remains an open question whether the J/$\psi$ is melt or not melt at RHIC energies. Finally, the STAR experiment has measured a smaller suppression at high transverse momentum (p$_{\rm T}\geq 5$ GeV/$c$) at mid-rapidity \cite{Star09b, Star12}  although the experimental errors remain large.

\subsection{The phase of deconfinement}

\begin{figure}
{\centering 
\resizebox*{0.48\columnwidth}{!}{\includegraphics{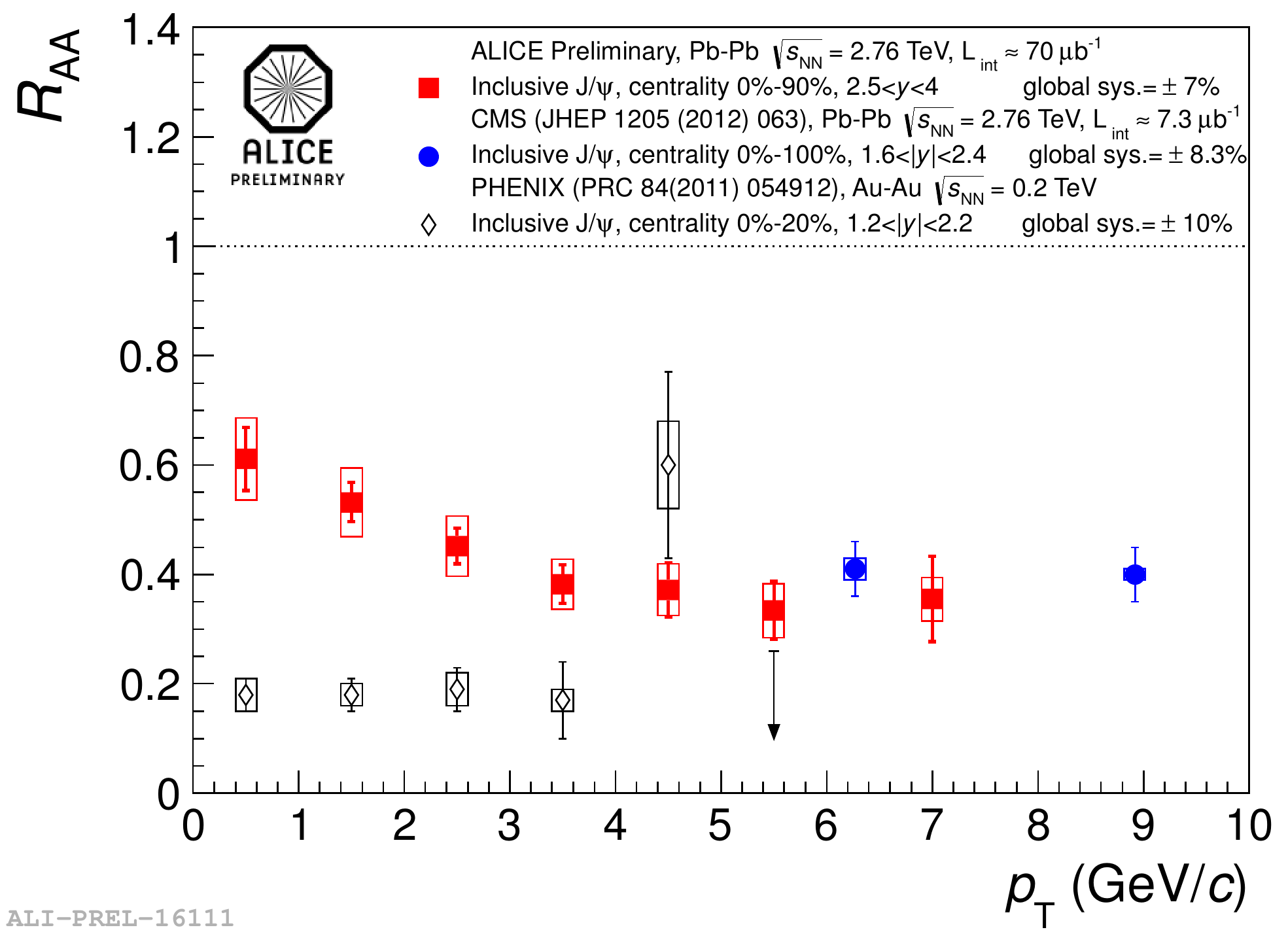}}
\resizebox*{0.48\columnwidth}{!}{\includegraphics{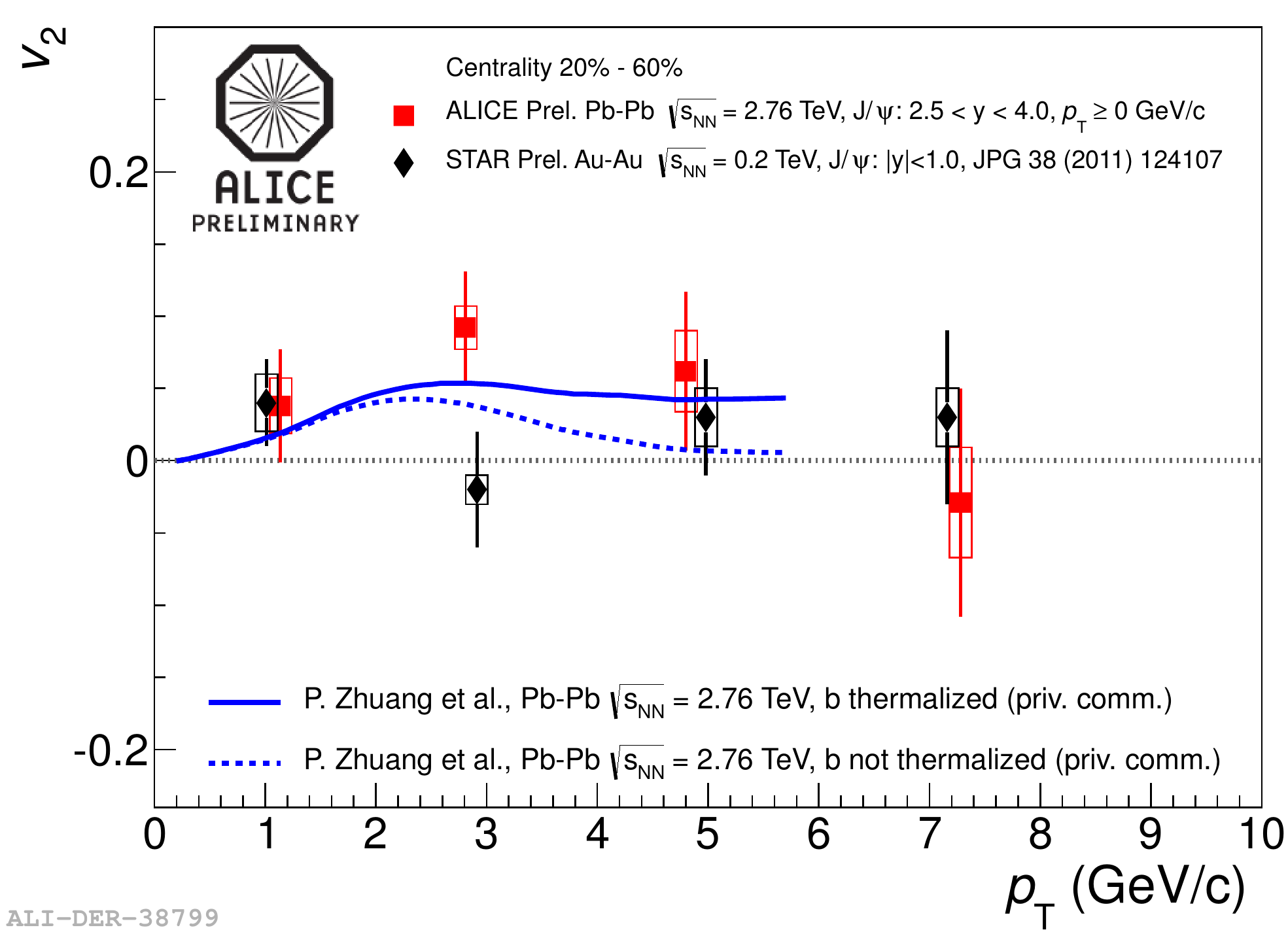}}
\par}
\caption{\label{1208-5601v2}
\emph{Left:  ALICE p$_{\rm T}$ dependence of the inclusive J/$\psi$ $R_{\rm AA}$ measured in Pb-Pb collisions at $\sqrt{s_{\rm NN}}$ = 2.76 TeV is compared to PHENIX and CMS measurements. Figure 2 in reference \cite{Suir12}.  Right: J/$\psi$ elliptic flow as a function of p$_{\rm T}$ in the centrality range 20-60\% and in the rapidity range $2.5 < y < 4$ (red squares). ALICE data are compared to STAR measurement performed in Au-Au collisions at $\sqrt{s_{\rm NN}}$=200 GeV and in the rapidity range $|y| < 1$ (black diamonds). ALICE data are compared with a transport model predictions. Figure 2 in \cite{Mass12}. }}
\end{figure}

At LHC energies, on average one J/$\psi$ particle is expected to be produced in every central Pb-Pb collision, together with about 50-100 $c$$\bar{c}$ quark pairs. As suggested in 1988 \cite{Svet88}, under these conditions the charm quark yield per unit of rapidity could be large enough to enhance the charmonium production in later phases of the hot QCD-matter dynamical evolution, in particular when the energy density is low enough to enable the charmonium bound state to be formed \cite{Brau00, Gran04, Andr11}. The ALICE collaboration reported the first measurement of the J/$\psi$ nuclear modification factor at LHC energies \cite{Alic12b}. Contrary to the expectations from $\Upsilon(2S)$ suppression, J/$\psi$ R$_{\rm AA}$ was found to be about 0.5 in the most central Pb-Pb collisions and does not exhibit a significant centrality dependence. In addition, the J/$\psi$ R$_{\rm AA}$ was found to be larger than that measured at RHIC. Contrary to RHIC observation, the J/$\psi$ R$_{\rm AA}$ is large at low  p$_{\rm T}$ and then decreases with increasing p$_{\rm T}$ \cite{Suir12} (see Fig. \ref{1208-5601v2} left). Finally, a hint of non-zero elliptic flow was also measured by the ALICE collaboration \cite{Mass12}  (see Fig. \ref{1208-5601v2} right). These experimental observations suggest that J/$\psi$ production at LHC energies is governed, for an important part, by charm quark recombination processes. The production of charmonium in latter stages of the QGP evolution would certainly be a direct probe of the deconfinement phase. 

\begin{figure}
{\centering 
\resizebox*{0.60\columnwidth}{!}{\includegraphics{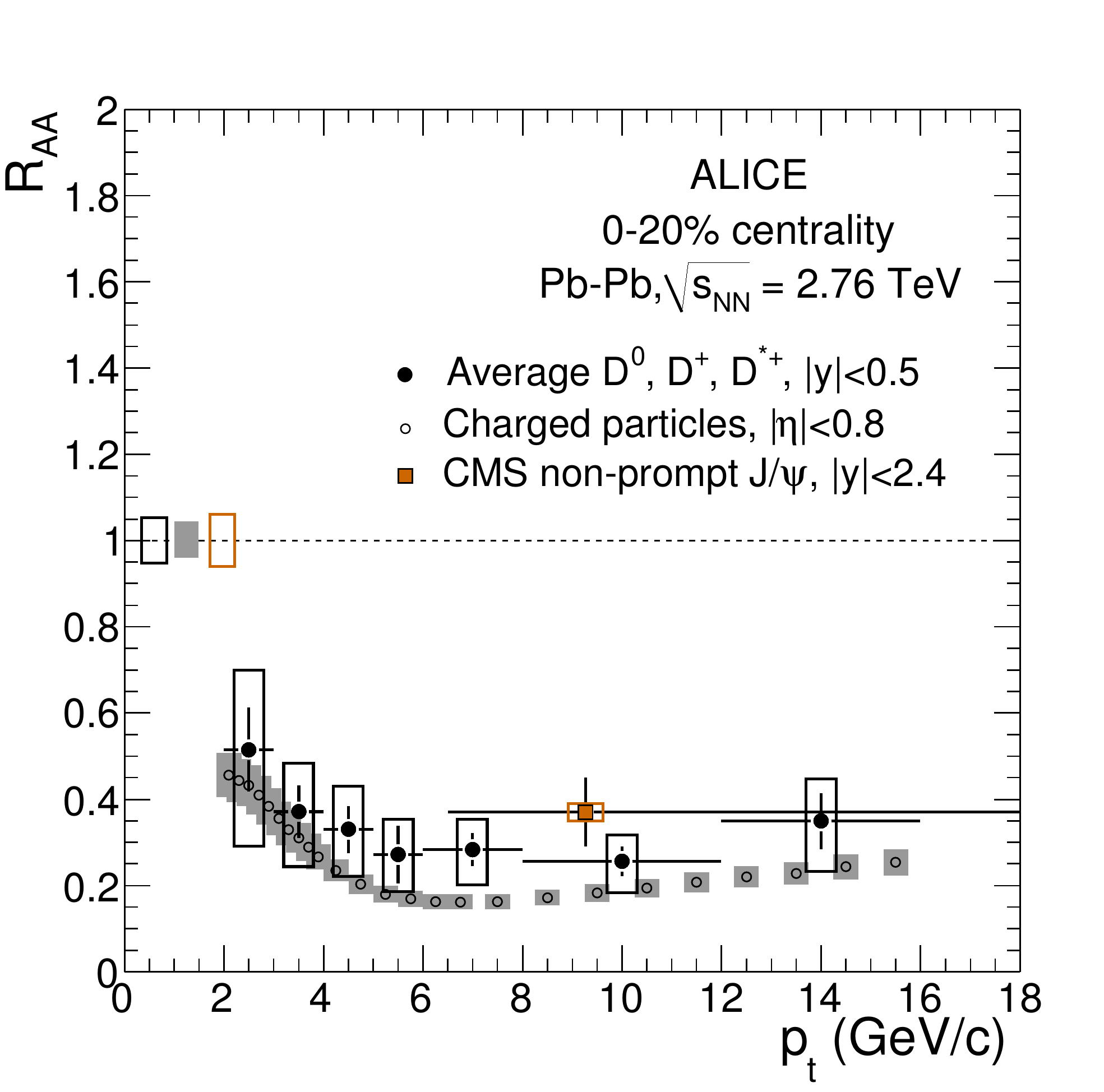}}
\par}
\caption{\label{zaida}
\emph{ Average $R_{\rm AA}$ of D mesons in the 0-20\% centrality class compared to the nuclear modification factors of charged particles and non-prompt J/$\psi$ from B decays measured by the CMS collaboration in the same centrality class. Figure 8 right in reference \cite{Alic12c}.}}
\end{figure}

Heavy-flavour hadrons, containing charm and beauty heavy quarks, are effective probes of the QGP. During the deconfined phase, low p$_{\rm T}$ heavy quarks will interact with the medium modifying its initial kinematical properties \cite{Hees07, Goss09} and, in the extreme scenario, they could become fully thermalised. The first ALICE results on the nuclear modification factor R$_{\rm AA}$ for charm hadrons in Pb-Pb collisions indicate strong in-medium energy loss for charm quarks. The D$^0$, D$^{+}$, and D$^{\star +}$ R$_{\rm AA}$, were measured for the first time as a function of transverse momentum and centrality. The suppression is almost as large as that observed for charged particles \cite{Alic12c, Cone12} (see Fig. \ref{zaida}).  A hint of non-zero flow of D hadrons was also measured \cite{Bian12}. High precision measurements of heavy-flavour hadrons at low p$_{\rm T}$ will remain an experimental challenge during the next 10 years at RHIC and the LHC.

\subsection{The opacity of the QGP}

\begin{figure}
{\centering 
\resizebox*{0.43\columnwidth}{!}{\includegraphics{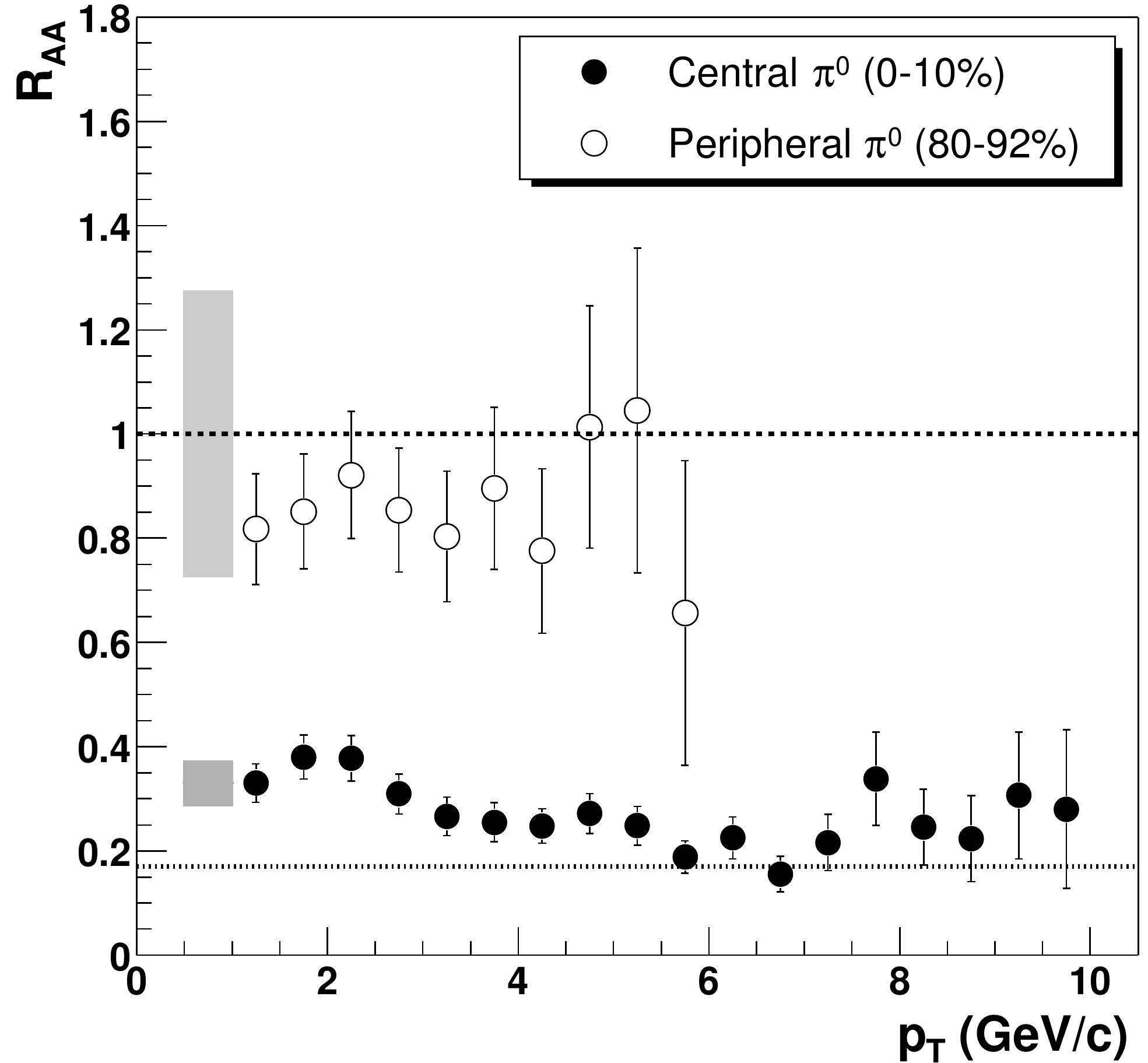}}
\resizebox*{0.50\columnwidth}{!}{\includegraphics{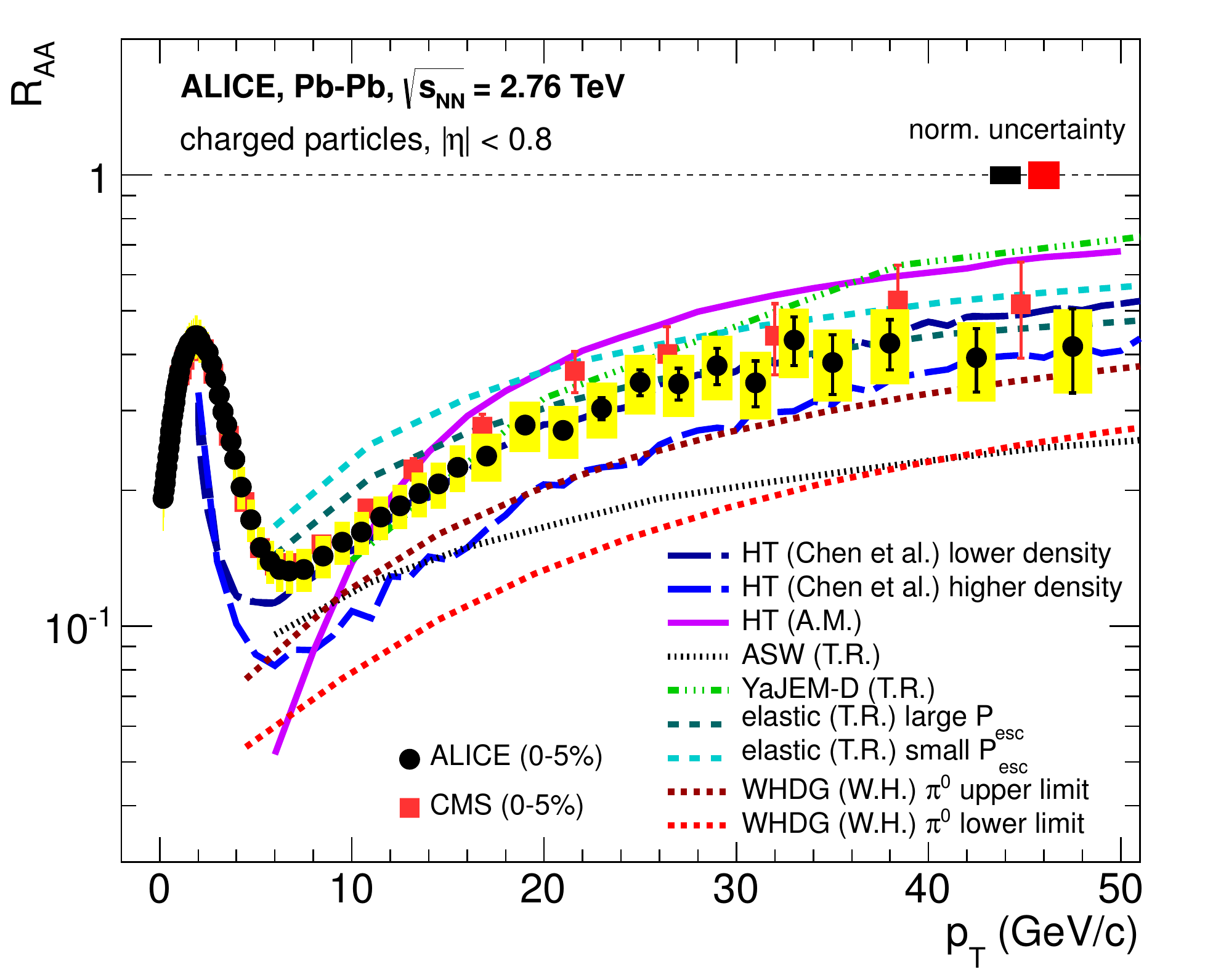}}

\par}
\caption{\label{raa}
\emph{Left: Nuclear modification factor R$_{\rm AA}$(p$_{\rm T}$) for $\pi^0$ in central 
(closed circles) and peripheral (open circles) Au-Au at $\sqrt{s_{_{\rm NN}}}$ 
= 200~GeV.  Figure 3 from reference \cite{Phen03b}. Right: $R_{\rm{AA}} $ of charged particles
 measured by ALICE in the most central Pb-Pb collisions (0-5\%) in comparison to results from CMS and model calculations. Figure 4 from reference \cite{Alic12e}}}
\end{figure}

Heavy ion collisions at LHC energies allowed to study for the first time the interaction between hard partons produced in the first stage of the hadronic collisions, with the QGP. The easiest experimental way to address this topic was via the study of high p$_{\rm T}$ particle yields and high p$_{\rm T}$ hadron-hadron correlations. High p$_{\rm T}$ particles are produced by the fragmentation of partons (quarks or gluons) in a time scale around $\tau_{\rm frag} \approx E/\Lambda_{\rm QCD} \times R$, where $E$ is the energy of the parton and $R$ is the typical size of a hadron ($\sim$ 1 fm). For energies above 5 GeV, the fragmentation time scale is about 20 fm/$c$. Therefore, in heavy ion collisions, partons are expected to fragment after traversing the QGP. One of the major discoveries at RHIC was the suppression of high p$_{\rm T}$ hadron R$_{\rm AA}$ and the quenching of back-to-back hadron correlations \cite{Star02,Phen03,Phen03b,Brah04,Star04b,Phen05}. This observation has been explained by the formation of a QGP drop where the initial hard partons interact losing a non negligible fraction of their initial energy (see section \ref{partonQGP}). QCD inspired models assuming gluon radiative energy loss in the QGP are in good agreement with the data \cite{Star05, Phen05}. On this topic the phenomenology is very rich and many experimental detailed studies have been performed \cite{Star11b, Star12b, Star12c, Phen08, Phen12b}.

The first LHC results on R$_{\rm AA}$  have confirmed RHIC results and extended the p$_{\rm T}$ ranges until values as high as 100 GeV/$c$ \cite{Alic10d, CMS12e, Alic12e}. The results indicate a strong suppression of charged particle production in Pb-Pb collisions and a characteristic centrality and p$_{\rm T}$ dependence of the nuclear modification factors. In the most central collisions, the R$_{\rm AA}$ is strongly suppressed (R$_{\rm AA}\approx$ 0.13) at p$_{\rm T}$ = 6-7 GeV/c. Above p$_{\rm T}$ = 7 GeV/c, there is a significant rise in the nuclear modification factor, which reaches R$_{\rm AA}\approx$  0.4 for p$_{\rm T}>$ 30 GeV/c (see Fig. \ref{raa}).  The latter is in good agreement with models based on radiative energy loss of gluons in QGP.

\begin{figure}
{\centering 
\resizebox*{0.98\columnwidth}{!}{\includegraphics{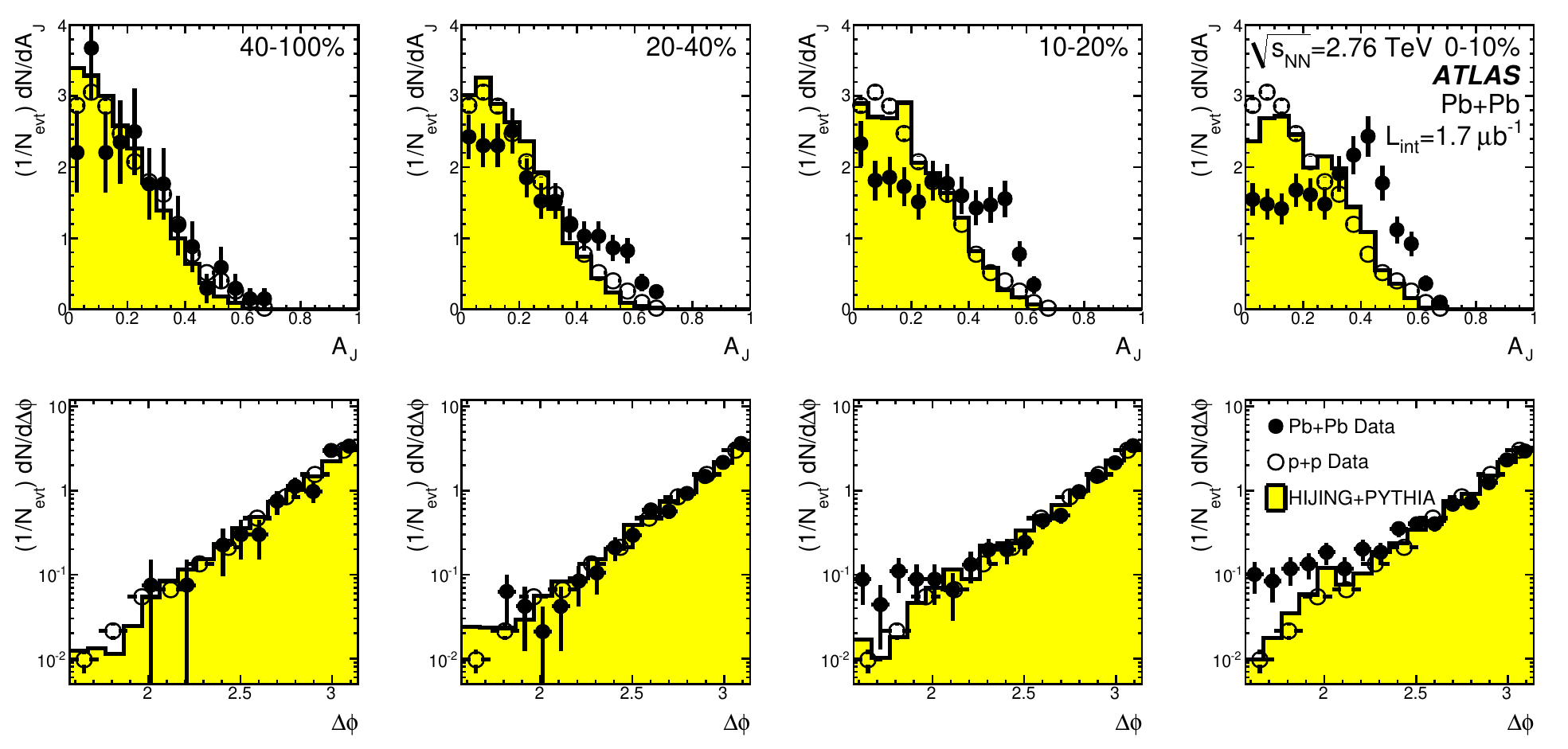}}
\par}
\caption{\label{dijets}
\emph{Top: dijet asymmetry distributions for data (points) and unquenched HIJING with superimposed PYTHIA dijets (solid yellow histograms), as a function of collision centrality (left to right from peripheral to central events). Proton-proton data from $\sqrt{s}$ = 7 TeV, analyzed with the same jet selection, is shown as open circles. Bottom: distribution of $\Delta \phi$, the azimuthal angle between the two jets, for data and HIJING+PYTHIA, also as a function of centrality. Figure 3 from reference \cite{Atla10}.}}
\end{figure}

At LHC the studies of jets in heavy ion collisions becomes possible.  The ATLAS collaboration presented the first results on jet reconstruction in Pb-Pb collisions at the LHC \cite{Atla10}. Jets were reconstructed up to  transverse energies of 100 GeV. An asymmetry, increasing with centrality, was observed between the transverse energies of the leading and second jets (see Fig. \ref{dijets}). This is an outstanding confirmation of the strong jet energy loss in a hot, dense medium, as it was inferred from the studies of the high p$_{\rm T}$ R$_{\rm AA}$ and hadron correlations at RHIC. Similar conclusions were obtained from the measurement performed by the CMS collaboration \cite{CMS11b}.

At LHC, the phenomenology on studies related to QCD energy loss is also very rich. Many measurements that are not described here have been performed, like hadron-hadron correlations \cite{Alic11b}, single jets \cite{CMS12f} and gamma-jets \cite{CMS12g}. In the next 10 years, high precision measurements will be performed on these channels and other more exotic ones, like Z-jet, will be studied.

The study of high  p$_{\rm T}$ R$_{\rm AA}$  of heavy flavour hadrons should shed light on the QCD energy loss mechanisms. According to QCD, the radiative energy loss of gluons should be larger than that of quarks. In addition, due to the dead cone effect \cite{Doks01}, heavy quark energy loss should be further reduced with respect to that of light quarks. Many studies were performed at RHIC, mainly via the semileptonic decay of heavy flavour hadrons. A strong suppression was observed but quantitative conclusions are not yet available. 
At the LHC, ALICE collaboration has measured the high p$_{\rm T}$ R$_{\rm AA}$ of D$^0$, D$^{+}$, and D$^{\star +}$  \cite{Alic12c, Cone12} and the high p$_{\rm T}$ R$_{\rm AA}$ of semi-muonic decay of heavy-flavours (charm and beauty) \cite{Alic12d}. The CMS collaboration has measured the high p$_{\rm T}$ R$_{\rm AA}$ of J/$\psi$ from beauty hadron decays. These results indicate strong in-medium energy loss for charm and beauty quarks, increasing towards the most central collisions. It seems that J/$\psi$ from beauty hadron decays are less suppressed than charm hadrons, but systematic uncertainties are still large. In the next 10 years, thanks to the upgrades of the LHC and RHIC experiments, higher precision measurements will become available.

\subsection{Other interesting measurements}

Among the huge amount of experimental results that have not been described in this section, I would like to quickly mention the following ones:

\begin{itemize}
\item  The measurement of electro-weak boson R$_{\rm AA}$, proposed by  \cite{Cone07}, has become possible at LHC. CMS and ATLAS collaboration has performed the first measurements at the LHC \cite{CMS11c, Atla12d, CMS12h}. These have been fundamental measurements and (unfortunately) the measured nuclear modification factor is compatible with unity, as it was expected.

\item The charged particle multiplicities measured in high-multiplicity pp collisions at LHC energies reach values that are of the same order as those measured in heavy-ion collisions at lower energies (e.g. they are well above the ones observed at RHIC for peripheral Cu-Cu collisions at 200 GeV \cite{Phob11}). Therefore, it is a valid question whether pp collisions also exhibit any kind of collective behaviour as seen in these heavy-ion collisions. An indication for this might be the observation of long range, near-side angular correlations (ridge) in pp collisions at 0.9, 2.36 and 7 TeV with charged particle
multiplicities above four times the mean multiplicity \cite{CMS10}.  Recently J/$\psi$ yields were measured for the first time in pp collisions as a function of the charged particle multiplicity density \cite{Alic12h}. The study of high multiplicity pp and p-A collisions will be an exciting topic in the next years.

\item Antimatter can efficiently be created in heavy ion collisions. STAR collaboration reported the first observation of the anti-helium-4 nucleus \cite{Star11c}.

\item Finally, ultra-peripheral heavy ion collisions at RHIC and at the LHC have become a powerful high luminosity photon beam. Many interesting measurements of vector meson $\rho$ \cite{Star08b, Star08, Star11c}, multi-pions \cite{Star10} or J/$\psi$  \cite{Phen09b, Alic12g} are being performed in both colliders.

\end{itemize}

\subsection{Caveat on cold nuclear matter effects}
This topic has not been addressed in the present proceedings. The study of cold nuclear matter effects in proton or deuteron induced collisions is of outstanding importance. Many of the interpretations of the experimental results given above can only be confirmed via the study of these collisions. At RHIC energies, deuteron induced collisions have been extensively studied. At LHC, the first run p-Pb has taken place beginning of 2013.

\section{Other lectures on QGP}

 The following references that will certainly complement the present lectures:
 
 \begin{itemize}

\item Lectures of Larry MacLerran, \emph{The Quark Gluon Plasma and The Color Glass Condensate: 4 Lectures} \cite{Lerr01}.

\item Lectures of Frithjof Karsch, \emph{Lattice Results on QCD Thermodynamics} \cite{Kars01b}.

\item Lectures of Jean-Paul Blaizot, \emph{Theory of the Quark-Gluon Plasma} \cite{Blai02}.

\item Lectures of Ulrich W. Heinz, \emph{Concepts of Heavy-Ion Physics} \cite{Hein04}.

\item Lectures of Anton Andronic and Peter Braun-Munzinger, \emph{Ultra relativistic nucleus-nucleus collisions and the quark-gluon plasma}, \cite{Andr04}.

\item Lectures of  Thomas Schaefer, \emph{Phase of QCD} \cite{Scha05}.

\item Lectures of Bernt M\"uller, \emph{From Quark-Gluon Plasma to the Perfect Liquid} \cite{Mull07}.

\item Lectures of Jean-Yves Ollitrault, \emph{Relativistic hydrodynamics for heavy-ion collisions} \cite{Olli07}.

\item Lectures of Tetsufumi Hirano, Naomi van der Kolk and Ante Bilandzic, \emph{Hydrodynamics and Flow} \cite{Hira08}.

\item Lectures of Carlos Salgado, \emph{Lectures on high-energy heavy-ion collisions at the LHC} \cite{Salg09}.

\item Article by Michael L. Miller, Klaus Reygers, Stephen J. Sanders, Peter Steinberg, \emph{Glauber Modeling in High Energy Nuclear Collisions} \cite{Mill07}.

\item Article by David d'Enterr\'{\i}a, \emph{Hard scattering cross sections at LHC in the Glauber approach: from pp to pA and AA collisions} \cite{Ente03}.

\item Article by Berndt Muller, Jurgen Schukraft and Bolek Wyslouch, \emph{First Results from Pb+Pb collisions at the LHC} \cite{Mull12}.

\item In French: Proceedings of Joliot-Curie School in 1998: \url{http://www.cenbg.in2p3.fr/heberge/EcoleJoliotCurie/coursJC/JOLIOT-CURIE%201998.pdf} \cite{Joli98}.

\item In French: Proceedings of Joliot-Curie School in 2005: \url{http://www.cenbg.in2p3.fr/heberge/EcoleJoliotCurie/coursJC/JOLIOT-CURIE%202005.pdf} \cite{Joli05}.

\item In French: My HDR (Habilitation \`a Diriger des Recherches) \emph{} \cite{Mart06}.

\end{itemize}

\section{Acknowledgements}
I would like to thank the organisers of the \href{http://ejc2011.sciencesconf.org/}{2011 Joliot-Curie School} for giving me the privilege to be one of the lecturers of this school which was devoted to the \emph{Physics at the femtometer scale} and, namely, commemorated the 30th edition of the school.

I would like to thank Bego\~na de la Cruz, Hugues Delagrange, Javier Martin and  Laure Massacrier for reading the manuscript, spotting many typos and making very interesting and fruitful comments.

I apologise to Navin Alahari, chairman of the Joliot Curie School, for the delay in getting ready these proceedings. He is the only person who knows how many electronic messages he sent to me as reminders of the different deadlines. Indeed he kindly agreed with several deadlines that I was not able to respect, except the last one.

\bibliography{EJC2011_Martinez}

\end{document}